\documentclass[aps,prd,preprint,12pt,superscriptaddress,nofootinbib,a4paper]{revtex4-1}
\usepackage[justification=raggedright,singlelinecheck=false]{caption}

\usepackage[utf8]{inputenc}
\usepackage{amsmath,amssymb}
\usepackage{graphicx}
\usepackage{hyperref}
\usepackage[table]{xcolor}
\usepackage{slashed}
\usepackage[]{subcaption}

\newcommand{\eqn}{equation}
\newcommand{\nn}{\nonumber}
\newcommand{\lb}{\left (}
\newcommand{\rb}{\right )}
\newcommand{\lam}{\lambda}
\newcommand{\pythia}{\textsc{Pythia 8.244}}
\newcommand{\mg}{\textsc{MG5\_aMC@NLO}}
\newcommand{\CM}{\textsc{CheckMATE}}
\newcommand{\T}{\ensuremath{\mathrm{T}}}
\newcommand{\met}{\ensuremath{E_{\mathrm{T}}^\mathrm{miss}}}
\newcommand{\spey}{\textsc{Spey}}
\newcommand{\thdma}{2HDMa}
\newcommand{\GeV}{{\ensuremath\rm GeV}}

\begin{document}
\rightline{RBI-ThPhys-2025-048}
\rightline{DESY-25-176}
\vspace{3mm}
\title{Constraining the Inert Doublet Model at the LHC }
\author{Jayita Lahiri}
\email{jayita.lahiri@desy.de}
\affiliation{II. Institut f{\"u}r Theoretische Physik,  Universit{\"a}t Hamburg, Luruper Chaussee 149, 22761 Hamburg, Germany}
\author{Tania Robens}
\email{trobens@irb.hr}
\affiliation{Ruder Boskovic Institute, Bijenicka cesta 54, 10000 Zagreb, Croatia}
\author{Krzysztof Rolbiecki}
\email{krolb@fuw.edu.pl}
\affiliation{Faculty of Physics,
University of Warsaw, ul.~Pasteura 5, PL-02-093 Warsaw, Poland}
\date{\today}



\begin{abstract}
In this work, we analyze experimental exclusion bounds that have been derived within a specific new physics realization, the two Higgs-doublet model with a pseudoscalar singlet (\thdma), and their application to a different model, the Inert Doublet Model (IDM), that features the same final state. In this context, we discuss the sensitivity of the ATLAS search for the \thdma{} in final states with leptons and missing energy. We demonstrate that, with cuts optimized for a specific model topology, other new physics scenarios with larger rates might yet escape detection. We also give an update on constraints from vector boson fusion production of the Standard Model-like scalar and subsequent invisible decay from full Run 2 data on the parameter space of the IDM, with a special emphasis on the off-shell region, as well as a search that specifically concentrates on soft lepton final states.
\end{abstract}
\maketitle
\section{Introduction}

The LHC has delivered a large amount of data, and besides an accurate measurement of properties of the Standard Model (SM) particles, another important goal is the investigation of models with additional matter content. Very often, models with different particle content and properties can lead to similar experimental signatures, posing the question whether the experimental collaborations should perform dedicated studies for each new physics scenario, or whether it is sufficient to limit the parameter space of such models using recasting tools. In case of the latter, new physics scenarios can be tested using Monte Carlo event samples within dedicated recasting frameworks as e.g.\ \CM~\cite{Drees:2013wra,Kim:2015wza,Dercks:2016npn,Desai:2021jsa,Lara:2025cpm}, while taking the SM background from an experimental publication.

In this work, we are therefore recasting several experimental searches using the full Run 2 data  within a specific new physics scenario that so far has not been directly tested by the LHC experiments. We consider the Inert Doublet Model (IDM) \cite{Deshpande:1977rw,Barbieri:2006dq,Cao:2007rm}, a two Higgs  doublet model that obeys an exact $\mathbb{Z}_2$ symmetry and therefore provides a viable dark matter candidate from the additional doublet. This model has been studied in great detail from the phenomenological perspective \cite{Swiezewska:2012eh,Gustafsson:2012aj,Arhrib:2013ela,Krawczyk:2013jta,Ilnicka:2015jba,Diaz:2015pyv,Belyaev:2016lok,Eiteneuer:2017hoh,Ilnicka:2018def,Kalinowski:2018ylg,Banerjee:2019luv,Tsai:2019eqi,Fabian:2020hny,Kalinowski:2020rmb,Basu:2020qoe,Banerjee:2021xdp,Banerjee:2021hal,Banerjee:2021anv,AbdusSalam:2022idz,Belyaev:2022wrn,Astros:2023gda,Justino:2024etz,Braathen:2024ckk,Abouabid:2025whn,Bal:2025nbu,Fan:2022dck} as well as in recasts (see e.g.\ \cite{Belanger:2015kga,Belyaev:2022wrn}), but currently no published experimental search exists that investigates this scenario.

One of the main production modes of the new physics particles in this model is Drell-Yan production of the novel particles. The new scalars do not couple to fermions, and therefore all decays are mediated via electroweak gauge bosons. A prominent signature is therefore an electroweak gauge boson and missing transverse momentum (MET) stemming from the dark matter particles in the final state, the mono-$Z$ signature. Such a signature is also a prominent signal in other models. We here focus on a search for the \thdma, the two Higgs doublet model with an additional pseudoscalar that couples to dark matter particles in the gauge eigenbasis~\cite{Ipek:2014gua,No:2015xqa,Goncalves:2016iyg,Bauer:2017ota,Tunney:2017yfp}. A major question we want to address is whether it is sufficient to recast searches for the 2HDMa to significantly constrain the IDM or whether dedicated searches for the latter are needed due to different production topologies in the two scenarios.

We additionally include other searches for new physics in our recasts of the IDM, in particular the search for invisibly decaying SM-like Higgs boson in the vector boson fusion (VBF) production mode~\cite{ATLAS:2022yvh} and the search for opposite sign same flavour (OSSF) lepton pairs with low transverse momentum~\cite{ATLAS:2019lng}. These searches provide constraints in complementary regions of the parameter space of the IDM. The latter search turns out to be sensitive to parameter points with low mass splittings, $\Delta m < 8$~GeV, filling the gap left by other searches, in particular~\cite{Lundstrom:2008ai} that uses a recast for the LEP supersymmetry search~\cite{EspiritoSanto:2003by} to cast bounds on the IDM.  

This paper is organized as follows. In Section~\ref{model} we briefly discuss the two models in comparison, namely, \thdma{} and IDM. We also discuss the constraints on the parameter space of the IDM. In Section~\ref{sec:simulation}, we detail our signal simulation to be used by \CM{} for recasting corresponding to three final states, namely $Z+$MET, VBF production of SM Higgs and soft-lepton final states. In Section~\ref{sec:results}, we present our results for the three aforementioned cases. Finally, we summarize our findings and conclude the discussion in Section~\ref{sec:summary}. Notes on implementation and validation of the searches in \CM{} can be found in the appendix.

\section{Models}
\label{model}

In this section we introduce the models, the \thdma{} and IDM. We briefly discuss Lagrangians and general features of the mass spectrum. Finally, we look at the production modes that can lead to similar final states, in particular the above-mentioned mono-$Z$ signature.

\subsection{\thdma}
\label{sec:2hdma}
The \thdma{} is a two Higgs doublet model that extends the scalar sector by an additional pseudoscalar field in the gauge eigenbasis that serves as a portal to a dark sector. The model has been introduced in Refs.~\cite{Ipek:2014gua,No:2015xqa,Goncalves:2016iyg,Bauer:2017ota,Tunney:2017yfp} and later proposed as a benchmark model for the LHC dark matter working group in \cite{Abe:2018bpo}. More recent work can also be found in \cite{Pani:2017qyd,Haisch:2018znb,Abe:2018emu,Haisch:2018hbm,Haisch:2018bby,Abe:2019wjw,Butterworth:2020vnb,Arcadi:2020gge,Argyropoulos:2021sav,Robens:2021lov,Arcadi:2022dmt,Arcadi:2022lpp,Haisch:2023rqs,Argyropoulos:2024yxo}. 

In the setup discussed here, the Yukawa structure of the model corresponds to the type 2 models, where up and down-type leptons couple to different doublets in the gauge eigenbasis (see e.g.~\cite{Branco:2011iw} for a review on two Higgs doublet models). Furthermore, we concentrate on a scenario with a fermionic dark matter candidate. This is the version that has been proposed in the LHC working group and has widely been investigated by the LHC experiments; recent Run 2 results can be found in \cite{CMS:2018zjv,ATLAS:2019wdu,CMS:2020ulv,ATLAS:2020yzc,ATLAS:2021yij,ATLAS:2021jbf,ATLAS:2021shl,ATLAS:2022ygn,ATLAS:2022znu,ATLAS:2023ild,ATLAS:2023rvb,ATLAS:2024vqf,ATLAS:2024fdw,ATLAS:2024kpy,ATLAS:2024vxm,ATLAS:2024rlu,CMS:2024rkj,CMS:2025xck}. 
 The scalar particle content of the model is given by two CP-even neutral scalars $h,\,H$, two CP-odd neutral scalars $a,\,A$ and a charged scalar $H^\pm$. In the context of \thdma, by convention, non-capitalized letters denote smaller masses. The dark matter candidate is fermionic and denoted by $\chi$.

The complete potential of the model consists of terms containing the standard 2HDM, the additional pseudoscalar part, and the coupling to dark matter (DM), respectively, and reads
\begin{align}
V_{\text{\thdma}}\, =&\, \mu_1 H_1^\dagger H_1 + \mu_2 H_2^\dagger H_2 
+ \lambda_1 (H_1^\dagger H_1)^2 + 
\lambda_2 (H_2^\dagger H_2)^2
+ \lambda_3 (H_1^\dagger H_1)(H_2^\dagger H_2) \nn\\
&+ \lambda_4 (H_1^\dagger H_2) (H_2^\dagger H_1)
 + \left[\mu_3 H_1^\dagger H_2 + {\lambda_5} (H_1^\dagger H_2)^2 + h.c.\right], \\
V_P \,=&\, 
  \frac{1}{2} m_P^2 P^2 + \lambda_{P_1} H_1^\dagger H_1 P^2 +
  \lambda_{P_2} H_2^\dagger H_2 P^2 +
  (i b_P H_1^\dagger H_2 P + h.c.),\label{eq:vp}\\
\mathcal{L}_\chi\,=&\,-i\,y_\chi P\bar{\chi}\gamma_5\,\chi.
\end{align}
After electroweak symmetry breaking, which proceeds analogously to the 2HDM, the model contains in total 14 free parameters, which we choose to be
\begin{\eqn}\label{eq:pars}
v,\,m_h,\,m_A,\,m_H,\,m_{H^\pm},\,m_a,\,m_\chi,\,\cos\lb \beta-\alpha\rb,\,\tan\beta,\,\sin\theta,\,y_\chi,\,\lam_3,\,\lam_{P_1},\,\lam_{P_2}.
\end{\eqn}
 Two of the above parameters, $v$ and $m_h$, are fixed by electroweak precision measurements and LHC measurements, respectively, leaving us with 12 free parameters. For a more detailed discussion of the model and the viable parameter space, please refer to \cite{Robens:2021lov, Arcadi:2022dmt, Arcadi:2022lpp, Argyropoulos:2024yxo}.

As it is in general difficult to display results in a multi-dimensional parameter space, it is customary to fix all but two of the above parameters, making it easier to display results in a two-dimensional plane. However, care must be taken to correctly interpret these results. In particular, different choices of the fixed parameters might lead to different exclusion bounds and predictions.

The model undergoes a large number of theoretical and experimental constraints, e.g.\ discussed in detail in \cite{Robens:2021lov}. In this work it was shown that, when going away from the standard scenarios where all but two parameters are fixed, it is not so obvious to apply direct stringent constraints on the model; instead, in most cases the allowed parameter space largely opens up. Notable exceptions are e.g. bounds in the $\lb m_{H^\pm};\,\tan\beta\rb$ plane from flavour physics or mass differences of the new scalars due to electroweak precision constraints. But in general, two-dimensional constraints are typically shown by the experimental collaborations, as e.g. in \cite{ATLAS:2022znu,ATLAS:2024vqf}. 

\begin{figure}[!hptb]
        \begin{center}
        \includegraphics[width=0.5\textwidth]{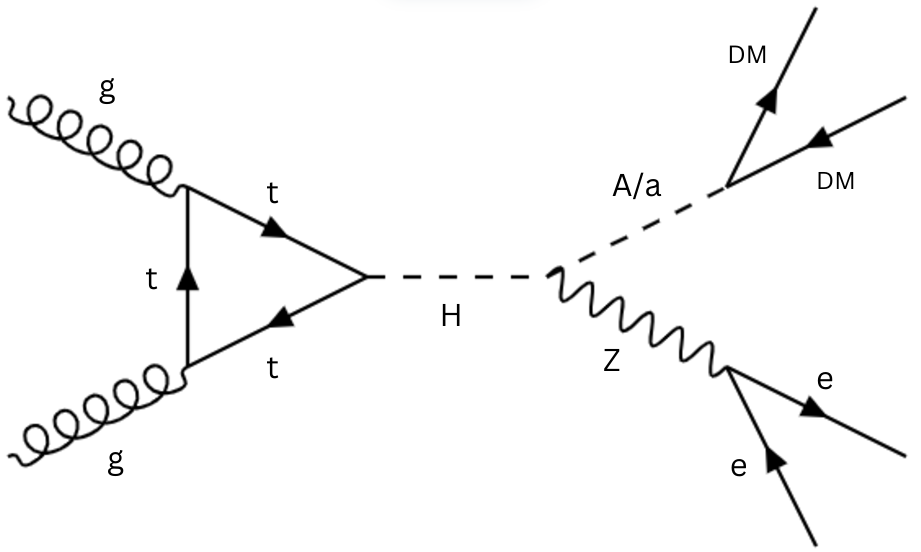}
        \caption{Production channel for $Z(\ell^+\ell^-)+E^\text{miss}_T$ final state in the \thdma.}
        \label{fig:feynman_2hdma}
        \end{center}
    \end{figure}

In this work, we use the experimental results presented in  Ref.~\cite{ATLAS:2021gcn} that impose constraints on the \thdma{} using the OSSF lepton pair plus missing energy final state. In Fig.~\ref{fig:feynman_2hdma}, we show the major production channel leading $Z(\ell^+\ell^-) + \text{MET}$ final state in the \thdma. The experimental search was implemented in \CM{} and in the following we examine its sensitivity to the IDM. For more details on the search and implementation in \CM{} we refer to Appendix~\ref{app:validation} and Ref.~\cite{Lara:2025cpm}.

\subsection{Inert Doublet Model}\label{sec:idm}

In the Inert Doublet Model~\cite{Deshpande:1977rw,Barbieri:2006dq,Cao:2007rm}, a second (dark) $SU(2)_L$ doublet  is added to the SM scalar sector. Furthermore, a discrete $\mathbb{Z}_2$ symmetry is imposed on the Lagrangian. The SM states and the dark sector (DS) transform under this $\mathbb{Z}_2$ symmetry as follows:
\begin{equation*}
    \text{SM}\,\rightarrow\,\text{SM};\;\text{DS}\,\rightarrow\,-\,\text{DS}.
\end{equation*}
The IDM differs from standard realisations of 2HDMs because of the unbroken $\mathbb{Z}_2$ symmetry, such that none of the beyond the SM (BSM) scalars mix with the SM-like states. The lightest particle of the so-called dark doublet is therefore stable and can, under certain conditions, provide a good dark matter candidate.\footnote{See e.g.\ Ref.~\cite{Kalinowski:2020rmb} for regions of parameter space where the IDM can saturate the observed DM relic density.}

The SM-like and the new doublets, denoted respectively $\Phi_1$ and $\Phi_2$, can be expanded as
\begin{align}
    \Phi_1=\begin{pmatrix}
      G^+\\
      \frac{1}{\sqrt{2}}(v+h+iG^0)
    \end{pmatrix}\,,
    \quad \text{and} \quad
    \Phi_2=\begin{pmatrix}
      H^+\\
      \frac{1}{\sqrt{2}}(H+iA)
    \end{pmatrix}\,,
\end{align}
where $h$ denotes the SM-like Higgs boson with a mass of 125 GeV and $H$, $A$, and $H^\pm$ are the physical mass eigenstates. Throughout this paper, in the IDM context, we will consider $H$ to be the lightest scalar and therefore the DM candidate\footnote{In principle, it is possible to choose $H^{\pm}$ as DM candidate. However, charged DM candidates are subject to strong constraints, see e.g. \cite{ParticleDataGroup:2024cfk}. 
}.  

The scalar potential of the IDM is given by 
\begin{align}
    V^{(0)}_\text{IDM} =&\ \mu_1^2 |\Phi_1|^2 + \mu_2^2 |\Phi_2|^2 + \frac{1}{2} \lambda_1 |\Phi_1|^4 + \frac{1}{2} \lambda_2 |\Phi_2|^4 + \lambda_3 |\Phi_1|^2 |\Phi_2|^2 + \lambda_4 |\Phi_1^\dagger \Phi_2|^2 \nonumber\\
    &+ \frac{1}{2} \lambda_5 \left[(\Phi_1^\dagger\Phi_2)^2+\text{h.c.}\right]\,. 
\end{align}
All parameters in this potential are taken to be real. After electroweak symmetry breaking, the model has in total 7 free parameters; we will comment on a typical parameter choice below.
Furthermore, our choice of $H$ as the DM candidate corresponds to taking $\lambda_5$ negative.
 We also note that either $H$ or $A$ can be the lightest $\mathbb{Z}_2$-odd neutral scalar and therefore dark matter depending on the sign of  $\lambda_5$.

Two parameters are then fixed by experimental findings, namely, the vacuum expectation value $v\,\sim\,246\,\GeV$ as well as $m_h\,\sim\,125\,\GeV$. The scalar sector of the IDM can be described in terms of five free parameters, which we choose to be
\begin{align}
    m_H,\ m_A,\ m_{H^\pm},\ \lambda_2,\ \text{and} \ \lambda_{345}\,. 
\end{align}

The model is subject to a large number of theoretical and experimental constraints. In this work, we follow the scan presented in \cite{Ilnicka:2015jba}, with further recent updates \cite{Ilnicka:2018def,Kalinowski:2018ylg,Dercks:2018wch,Kalinowski:2020rmb,Braathen:2024ckk,Bal:2025nbu}.\footnote{We do not apply the two-loop constraints on the scalar couplings presented in Ref.~\cite{Braathen:2024ckk}, but choose to use the leading order values for constraints.} We check for constraints on the scalar potential from boundedness-from-below, perturbativity, perturbative unitarity and electroweak precision observables. We have evaluated these quantities with the \textsc{2HDMC} computer program \cite{Eriksson:2009ws,Eriksson:2010zzb}. The electroweak precision constraints are checked using the oblique parameters, namely $S,\,T,\,U$ \cite{Altarelli:1990zd, Peskin:1990zt,Peskin:1991sw,Maksymyk:1993zm} and comparing to their PDG values \cite{ParticleDataGroup:2024cfk}. We also make sure that the parameter point rests in the inert vacuum \cite{Ginzburg:2010wa}. We furthermore test for the most recent bound on 125~GeV Higgs decays to invisible particles assuming on-shell decay, $\text{BR}\lb h\,\rightarrow\,\text{invisible}\rb\,\leq\,0.11$ \cite{ATLAS:2023tkt}.
The model parameter space receives strong constraints from the upper limit on Higgs invisible branching ratio~\cite{CMS:2022qva}, when $m_H \lesssim \frac{m_h}{2}$. This puts a stringent bound on the Higgs-DM portal coupling $\lambda_{345}$ in that mass region.

We also considered dark matter constraints. We calculate relic density for our parameter points using \textsc{micrOMEGAs} \cite{Belanger:2018ccd} and demand agreement with the upper bound for relic density as measured by the Planck experiment \cite{Planck:2018vyg}. Note that we allow for points which produce less than the observed relic density. Direct detection constraints are taken from the LUX-ZEPLIN collaboration \cite{LZ:2024zvo} and {{put stringent bounds on this model. The Higgs resonance region ($m_H \approx \frac{m_h}{2}$) is interesting as it is possible to satisfy direct detection constraints while at the same time finding points that render exact relic density as measured by the Planck collaboration.}} As mentioned earlier, we have also allowed regions which are under-relic and for such points the direct detection cross section is scaled according to the following relation:
 \begin{equation}
     \sigma_{\textrm{DD}}^{\textrm{scaled}} =  \sigma_{\textrm{DD}} \frac{\Omega}{\Omega_{\textrm{obs}}},
 \end{equation}
 where $\Omega_{\textrm{obs}}$ is the observed relic density. The scaled direct detection cross section $\sigma_{\textrm{DD}}^{\textrm{scaled}}$ is compared with the upper limit on DM-nucleon scattering cross section from LUX-ZEPLIN.

Finally, we also make sure the mass hierarchy obeys the constraints from a recast of a SUSY search at LEP reinterpreted within the IDM \cite{Lundstrom:2008ai}, and that the particles would not contribute to the decay widths of the electroweak gauge bosons. More details on the actual scan procedure applied can be found in refs. \cite{Ilnicka:2015jba,Ilnicka:2018def,Kalinowski:2018ylg,Dercks:2018wch,Kalinowski:2020rmb,Braathen:2024ckk}.

\begin{center}
    \begin{figure}
        \begin{center}
        \includegraphics[width=0.6\textwidth]{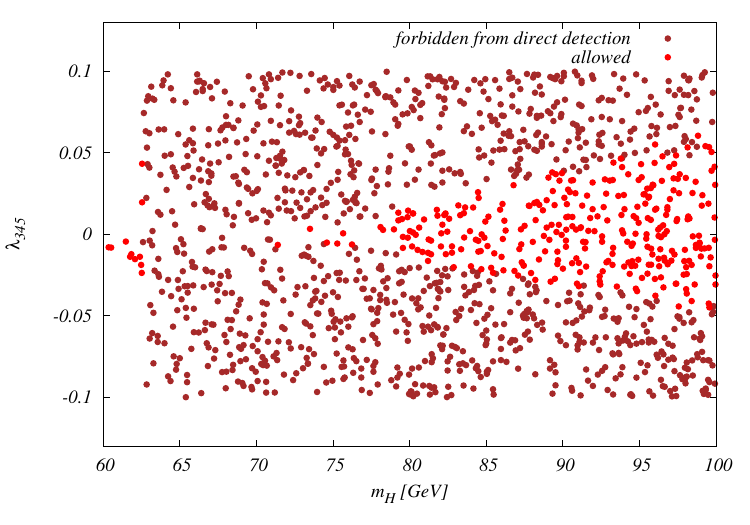}
        \caption{\label{fig:dirdet} Bounds from direct detection in the $\lb m_H, \lam_{345}\rb$ plane, including points forbidden by direct detection from the LUX-ZEPLIN experiment \cite{LZ:2024zvo} {\sl (brown)} as well as points allowed by all constraints {\sl (red)}. 
        }
        \end{center}
    \end{figure}
\end{center}

The impact of the aforementioned constraints on the parameter space of the model has been discussed in great detail in \cite{Ilnicka:2015jba,Kalinowski:2020rmb,Braathen:2024ckk} and is therefore not repeated here. 
As stated previously, it is worth noting that displaying bounds in a two-dimensional plane is not always straightforward, as quite often various free parameters enter the respective constraints. Counter examples of this can be found in the references mentioned above. For demonstration we show, in Fig. \ref{fig:dirdet}, the most recent constraint from direct detection in the $\lb m_H; \lam_{345}\rb$ plane. This is in particular important as these bounds severely constrain $\lam_{345}$ that governs the $h\,H\,H$ coupling. Other examples for distinct two-dimensional bounds are e.g.\ limits on mass differences from electroweak precision data as well as perturbative constraints in this model. 

If not mentioned otherwise, all parameter points that appear in the scans below have been required to fulfill all theoretical and experimental constraints discussed above.

\begin{figure}[!hptb]
\begin{center}
\subcaptionbox{\label{fig:domproc}}{\includegraphics[width=0.5\textwidth]{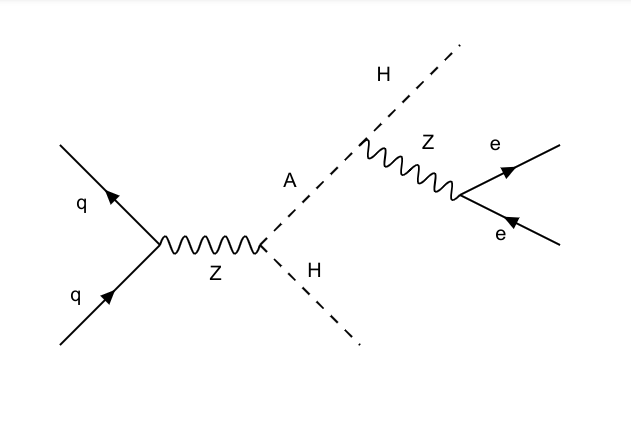}} \vspace{1cm} \\
\subcaptionbox{\label{fig:higgsstrahl}}{\includegraphics[width=0.45\textwidth]{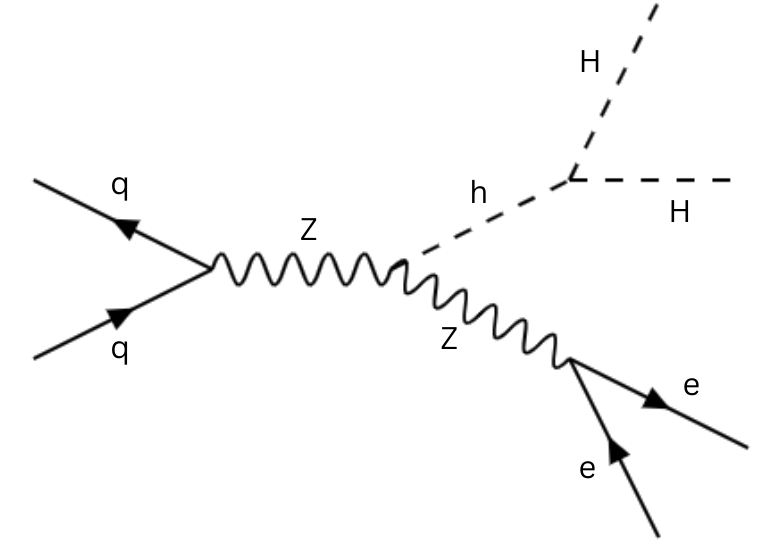}}\hfill
\subcaptionbox{\label{fig:chargedhiggs}}{\includegraphics[width=0.45\textwidth]{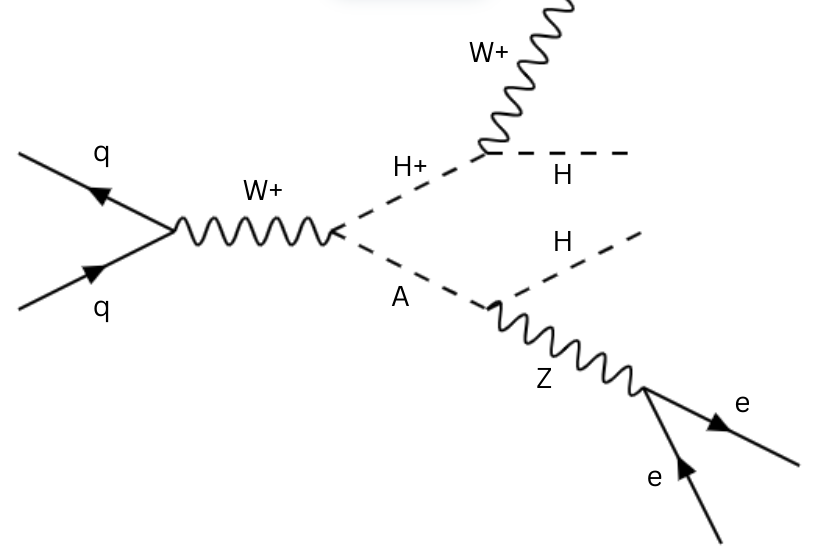}}
\end{center}
 \caption{Leading {\sl (top)} and sub-leading {\sl (bottom left)} and {\sl (bottom right)} production channels for $Z(\ell^+\ell^-$)+MET final state in the IDM. 
}
        \label{fig:feynman_IDM}
\end{figure}

Regarding the final state investigated here, in Fig.~\ref{fig:feynman_IDM} we show the leading and sub-leading production channels leading to the $Z(\ell^+\ell^-)+E^\text{miss}_T + X$ final state in the IDM. There are three Drell-Yan production modes. The leading one, diagram (a) in Fig.~\ref{fig:feynman_IDM}, is the associated production of $H$ and $A$, followed by $A\to H Z$ (here $H$ is a DM candidate) or $A\to H \ell^+ \ell^-$ if $m_A - m_H \lesssim m_Z$.\footnote{{Indeed, the off-shell effects will be non-negligible even if the mass splitting exceeds $m_Z$ by several GeV. These effects are taken into account in our Monte Carlo (MC) simulations.} } The Higgs-Strahlung diagram, Fig.~\ref{fig:feynman_IDM}(b), provides a subleading contribution, if the decay mode $h\to HH$ is open. The corresponding coupling $\lam_{345}$ is however subject to strong constraints from the invisible Higgs boson branching ratio. Finally, there is an $s$-channel $W$ boson exchange, diagram (c) in Fig.~\ref{fig:feynman_IDM}, leading to the associated production of $H^\pm$ and $A$, followed by $H^\pm\to W^\pm H$ and $A\to Z H$. For the hadronic $W$ decay this process results in additional jets in the final state. We discuss this case in more detail in the next section in the context of experimental selections.  
\section{Simulations}
\label{sec:simulation}
In this section we present our pipeline for the production of Monte Carlo events. We start with presenting common features and in the following we discuss details for particular processes.

The regions of parameter space that will be sensitive to the searches investigated here depend on both couplings and kinematics. In this sense, the region of the space that are constrained do not necessarily overlap.  For each of the three investigated production modes and final states, we therefore optimized the input parameters of our scans accordingly in order to cover the largest region of sensitive parameter space.

Events are generated using the \mg5{} framework~\cite{Alwall:2014hca,Alwall:2007fs,Alwall:2008qv} at the parton level using \textsc{NNPDF23LO} PDF set~\cite{Ball:2012cx,Buckley:2014ana}. For scenarios studied in this work, we used publicly available UFO~\cite{Darme:2023jdn} models for \thdma{}~\cite{ulirep, Bauer:2017ota} and the IDM~\cite{idmrep,Goudelis:2013uca}.  The events are then interfaced to \pythia~\cite{Sjostrand:2014zea} for parton showering and hadronization. Jet clustering is performed using \textsc{fastjet 3.3.4}~\cite{Cacciari:2011ma,Cacciari:2005hq,Cacciari:2008gp}. The resulting events are then passed to \CM{}~\cite{Kim:2015wza,Dercks:2016npn,Lara:2025cpm} to test against ATLAS searches: for the invisible Higgs decays in vector boson fusion ~\cite{ATLAS:2022yvh}, for supersymmetric winos and higgsinos in the final states with soft leptons \cite{ATLAS:2019lng} and for the search for events with dilepton and missing transverse energy \cite{ATLAS:2021gcn}. \CM{} is a universal tool for recasting of LHC searches within arbitrary New Physics models. It uses the fast detector simulation framework \textsc{Delphes}~\cite{deFavereau:2013fsa} with a customized ATLAS detector card and additional built-in tuning for a more accurate reproduction of experimental efficiencies.

The statistical evaluation of multibin signal regions is performed using \spey{} program~\cite{Araz:2023bwx} and the package \textsc{Pyhf}~\cite{Cowan:2010js,pyhf_joss,pyhf}, which is a Python implementation of the \textsc{HistFactory} specification for binned statistical models~\cite{Cranmer:1456844,Baak:2014wma}. In the soft lepton search, ATLAS SUSY-2018-16, we use the full likelihood model provided by ATLAS, see Ref.~\cite{ATL-PHYS-PUB-2019-029} for details. The shape-fit in the mono-$Z$~\cite{ATLAS:2021gcn} search is performed following a simplified approach detailed in Ref.~\cite{ATL-PHYS-PUB-2021-038}. More details regarding the implementation and validation of statistical models in \CM{} can be found in Ref.~\cite{Lara:2025cpm}.  

The results of evaluation in the case of signal region bins are reported in \CM{} using an $r$ parameter, which compares the observed number of events from the Monte Carlo simulation to the observed 95\% CL reported by an experiment. It is defined conservatively\footnote{This definition increases the exclusion threshold by the combined $2-\sigma$ uncertainty in the Monte Carlo and theory predictions. In particular it avoids apparent over exclusion due to random fluctuations in the MC simulation.} as (see e.g.\ \cite{Dercks:2016npn}):
\begin{equation}\label{eq:rdef}
    r = \frac{S-1.64\cdot \Delta S}{S95}
\end{equation}
where $S$ is the number of signal events recorded in a bin, $\Delta S$ is the uncertainty on $S$, and $S95$ is the 95\% CL limit on the number of events reported by an experiment. The nominator is the one-sided 95\% CL lower limit on signal events as calculated by \CM{}.\footnote{If a two-sided 95\% CL was used, 1.64 would be replaced by 1.96 in the above expression.}  $\Delta S$ is typically dominated by the MC uncertainty and can be neglected if the number of simulated events is sufficiently large. Hence, a parameter point is excluded if $r > 1$, i.e. if the lower limit on the number of MC-generated events exceeds the upper limit from experiment for the specific point.

In case of the multibin searches the final evaluation is decided using the upper limit on the combined production and decay signal strength $\mu$ and a parameter point is excluded (allowed) if $\mu < 1$ ($ \mu \ge 1 $).
In this approach, $\mu = 1$ is the theoretically predicted signal for a given parameter point. It is easy to see that, neglecting the MC uncertainty $\Delta S$, we have $\mu = 1/r$.

In the following, we list the different search channels investigated in this work. Details on the validation can be found in Appendix \ref{app:validation}.

\subsection{$Z$ and missing energy}

\noindent
Here we concentrate on the signal events with a leptonically decaying $Z$ boson and missing transverse energy, leading to the dilepton and missing transverse energy signature
\begin{equation}\label{eq:zmetsig}
    p\,p\,\rightarrow\,\ell^+\,\ell^-\,+\,\met.
\end{equation}
The original ATLAS study~\cite{ATLAS:2021gcn} aimed at constraining the invisible decays the Higgs boson produced in association with a $Z$ boson as well as constraining the benchmark DM model \thdma. As discussed in the introduction a similar final state, albeit with different topology, occurs in the IDM.

\begin{figure}[!hptb]
\begin{center}
\includegraphics[width=0.49\textwidth]{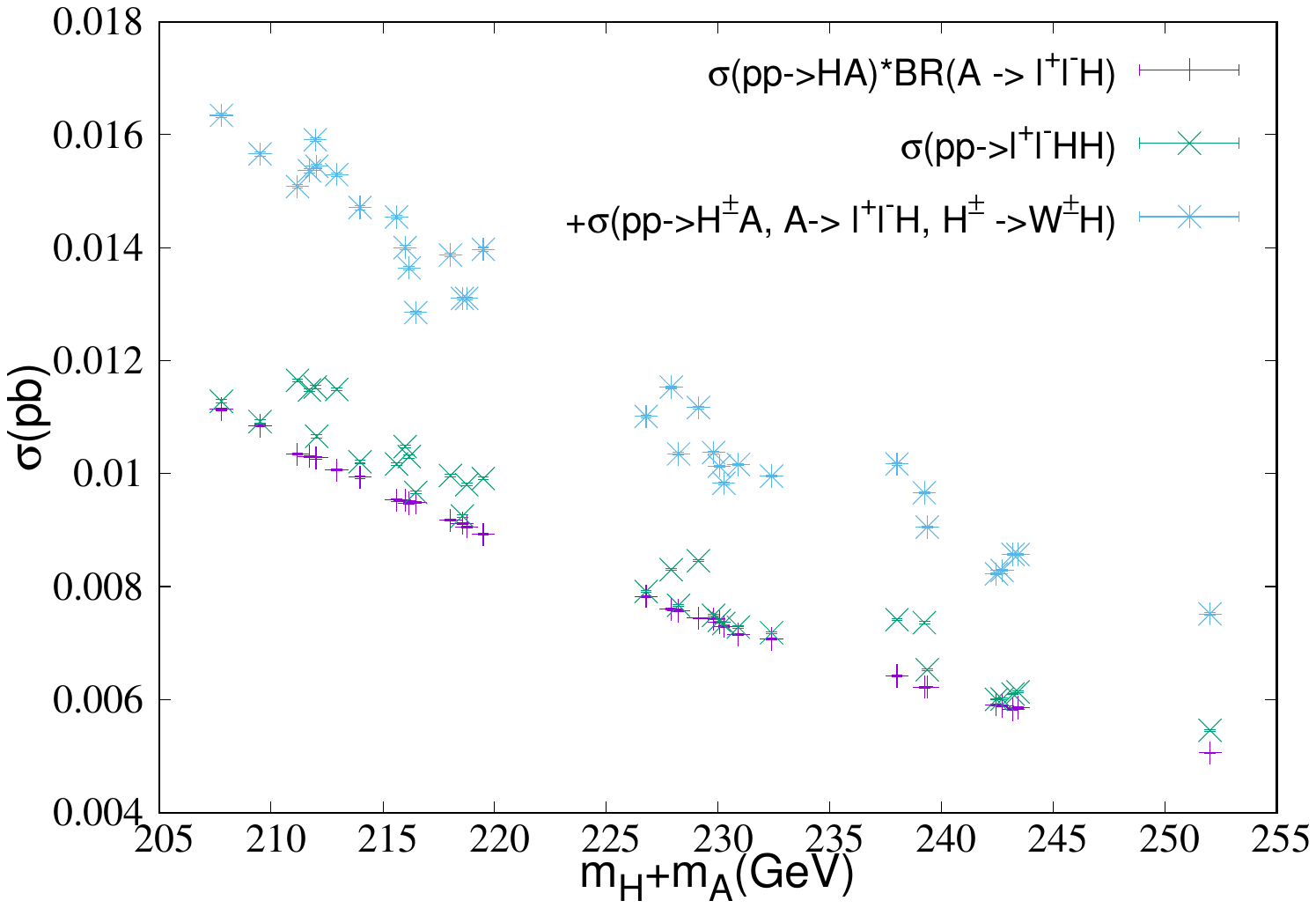}
\includegraphics[width=0.49\textwidth]{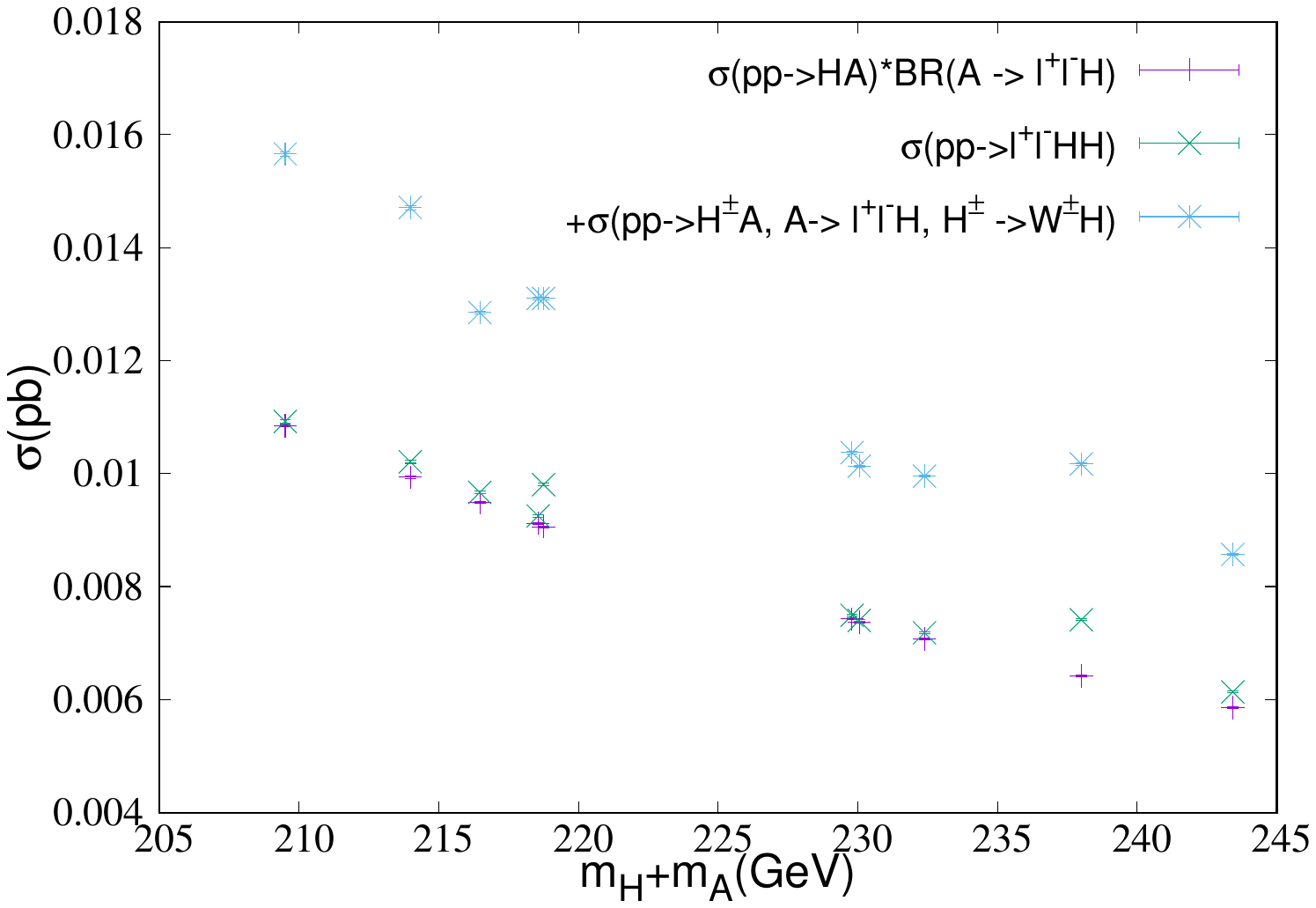}
\caption{{\sl Left:} Contribution from various production processes to the $Z(\ell^+\ell^-$)+MET final state. The purple points {\sl (bar)} denote the Drell-Yan cross section, i.e.\ on-shell $HA$ production and subsequent decay, the green points {\sl (cross)} refer to the cross section of the full process $pp\rightarrow \ell^+\ell^-HH$. In the cyan points {\sl (star)}, the process involving charged scalars was added to the target process $pp\rightarrow \ell^+\ell^-HH$. {\sl Right:} a subset of the points shown in the plot on the left panel, that satisfy the DM constraints.
}
\label{fig:crosssec}
\end{center}
\end{figure}

The following set of cuts is used for event selection~\cite{ATLAS:2021gcn}:
\begin{itemize}
\item $p_T$ of the leading and subleading leptons $> 30, 20$~GeV respectively.
\item The invariant mass of the di-lepton pair should correspond to the $Z$-boson mass within a certain window, 76 GeV $< m_{\ell\ell} <$ 106~GeV.
\item $\met > 90$~GeV.
\item Distance between the lepton pair $\Delta R_{\ell\ell} <$ 1.8.
\end{itemize}
In our MC simulations, we consider the following production processes corresponding to the mono-$Z$ final state in IDM (see figure \ref{fig:feynman_IDM}):
\begin{align}
    & pp \to A H,\quad A \to \ell^+ \ell^- H & \label{eq:domproc}\\
    & pp \to h Z(\ell^+ \ell^-), \quad h \to H H & \label{eq:higgsstrahlung}\\
    & pp \to A H^\pm\quad A \to \ell^+ \ell^- H, \quad H^\pm \to W^\pm H,& \label{eq:ws}
\end{align}

\noindent
where $\ell = e,\ \mu$, see Fig.~\ref{fig:feynman_IDM}. If kinematically allowed, the major contribution to the mono-$Z$ final state comes from the on-shell production of the $HA$ pair, where $A$ further decays into $Z(\ell^+\ell^-)\,H$ Eq.~\eqref{eq:domproc}. However, the sub-leading contributions are also important in our analysis, as we will discuss in more detail below. The Higgs-Strahlung process Eq.~\eqref{eq:higgsstrahlung}, with the Higgs boson decaying into a pair of dark matter particles,  can become important when the Higgs-portal coupling $\lambda_{345}$ is non-negligible. On the other hand, the associated production of $AH^{\pm}$ Eq.~\eqref{eq:ws}, which is proportional the SM gauge coupling, also features a substantial production cross section depending on a specific point in the parameter space. If the $W$ boson produced in the process decays leptonically and the lepton evades detection, this process will contribute to the di-lepton+MET final state. As the search we investigate does not apply a jet veto,
in cases where the $W$ decays hadronically, this process can contribute significantly to the target signature \eqref{eq:zmetsig}. We compare the cross sections for each of these processes in Fig.~\ref{fig:crosssec}\footnote{Note that in the left panel of that figure, we partially ignore bounds from direct detection to better exemplify effects of Higgs-strahlung contributions. In the right panel we only show points that are allowed by all constraints.}. The total contribution for the two sub-leading processes Eq.~\eqref{eq:higgsstrahlung} and \eqref{eq:ws} can render up to 25-30\% of the total rate, compared with the on-shell $HA$ production. However, one should note that, not all points in Fig.~\ref{fig:crosssec} are allowed by the dark matter constraints. Such constraints particularly disfavor large portal couplings namely, $\lambda_{345}$. Since the same coupling drives the Higgs-Strahlung process as well, the dark matter constraints put strongest restriction on the contribution to the total cross-section through Higgs-Strahlung. We show the subset of points from Fig.~\ref{fig:crosssec} that satisfy the dark matter constraints in the right panel. 

Note that we do not explicitly require an intermediate $Z$ boson in the decays of the heavy scalar. This allows for a correct inclusion of kinematic effects due to the finite width of the $Z$ boson for parameter points where $m_A - (m_Z + m_H) < 10$~GeV or even when the $Z$ boson cannot go on-shell. On the other hand the $W$ boson in the decay of the charged scalar is always on-shell and far enough from the threshold. The $W$ decay gives either an additional lepton, in which case if the lepton is detected the event is vetoed, or additional jets which are allowed by the experimental selections. The process with a charged scalar contributes around $10$--$20$\% to the total number of selected events.

\subsection{Higgs boson production in vector boson fusion}
In Ref.~\cite{Dercks:2018wch}, constrains on the IDM were derived using a CMS search of an invisibly decaying Higgs boson produced through vector boson fusion~\cite{CMS:2018yfx} and an ATLAS search for dark matter candidates~\cite{ATLAS:2017dnw} in the monojet channel. 
 We update the former result using a search from ATLAS~\cite{ATLAS:2022yvh} based on the full Run 2 luminosity. Since the monojet search showed an inferior sensitivity to the IDM we do not include monojet searches in our analysis.

We consider the ATLAS search for invisible decays of the Higgs boson produced in vector boson fusion events, EXOT-2020-11~\cite{ATLAS:2022yvh}. The final state consists of two forward jets consistent with a vector boson fusion signature and missing transverse momentum. There is one multibin signal region divided into 16 bins according to a number of jets, invariant mass of the leading jets, $m_{jj}$, and their angular separation. In the analysis we only use the best signal region, as explained in the validation notes provided in Appendix~\ref{sec:vbf_vali}. 

We consider the following production process at the LHC:
\begin{equation}
    pp \to H\,H\,j\,j\,.\label{eq:vbf}
\end{equation}
The following generator-level cuts are applied in line with the signal selection of Ref.~\cite{ATLAS:2022yvh} in order to increase production efficiency:
\begin{itemize}
    \item leading jets transverse momentum, $p_{T,j1}>70$~GeV,  $p_{T,j2}>40$~GeV,
    \item invariant mass of two leading jets, $m_{jj} > 500$~GeV,
    \item separation of leading jets: $\Delta R > 3$.
\end{itemize}

Following the approach of Ref.~\cite{Dercks:2018wch} we also include a $K$-factor of $1.67$ and compare the results with and without the $K$-factor, see the Appendix~\ref{sec:vbf_vali} for details. {The cross sections for process~\eqref{eq:vbf} range from $\mathcal{O}(1\ \mathrm{fb})$ to $\mathcal{O}(1\ \mathrm{pb})$ in the scanned section of the parameter space.}

\subsection{Soft leptons}
Another interesting corner of the IDM parameter space is a region where mass differences, in particular between $A$ and $H$ bosons, are small. In such a case the $Z$-boson mediated decay of the heavy scalar,
\begin{equation}\label{eq:AllH}
    A \to Z^* H \to \ell^+ \ell^- H,
\end{equation}
will result in a opposite-sign same-flavour pair of soft leptons and missing energy. The invariant mass distribution of the lepton pair will have an end-point at $m_A - m_H$. This signature is similar to final states arising in supersymmetric models with compressed mass spectra, that was extensively studied by ATLAS~\cite{ATLAS:2019lng,ATLAS:2021moa} and CMS~\cite{CMS:2021edw}. As already mentioned, LEP excluded parameter points with $m_A - m_H >  8$~GeV. On the other hand even smaller mass difference will result in leptons having a very low transverse momentum. In order to detect such leptons, one needs an additional transverse boost which is provided by hard initial state radiation (ISR) jets. While originally optimized for SUSY models this search offers sensitivity to parameter points in the Inert Doublet Model, when the mass difference between scalars is small. In such a case the decay mode \eqref{eq:AllH}
is similar to the decay of the second lightest neutralino to the lightest neutralino (missing transverse energy) and a OSSF lepton pair:
\begin{equation}
    \tilde{\chi}_2^0 \to \ell^+ \ell^-  \tilde{\chi}_1^0
\end{equation}

In order to  test LHC constraints on the IDM with small mass difference $m_A - m_H < 8$~GeV we simulate, using \mg{}, events as follows: 
\begin{equation}
    pp \to A\ \mathcal{H}\ j \label{eq:prodsoft}
\end{equation}
where $\mathcal{H} =\{ A, H, H^+, H^-\}$ and $A$ is required to decay as in \eqref{eq:AllH}.\footnote{The event selection allows for additional jets in the final state, however events with an additional lepton (e.g.\ from $H^\pm$ decay) will be vetoed. } The ISR jet is required to have $p_T > 70$~GeV. This generator level cut is motivated by the selection in the soft lepton ATLAS search~\cite{ATLAS:2019lng}. In our setup the cross section for the process \eqref{eq:prodsoft} for phenomenologically interesting (within LHC sensitivity) points is in the range 10--20~fb.

\section{Results for the IDM}
\label{sec:results}
\noindent
Having discussed three different final states and corresponding selection criteria, we present the results of recasting of LHC data on each final state on to IDM parameter space. 

\subsection{$Z$ and missing energy}

We first discuss the impact of the $Z(\ell^+\ell^-) + \met$ search on the parameter space of the IDM. As mentioned earlier, we have scanned the parameter space against all the constraints mentioned in section~\ref{sec:idm}.

   \begin{figure}[!hptb]
        \begin{center} 
        \includegraphics[width=0.48\textwidth]{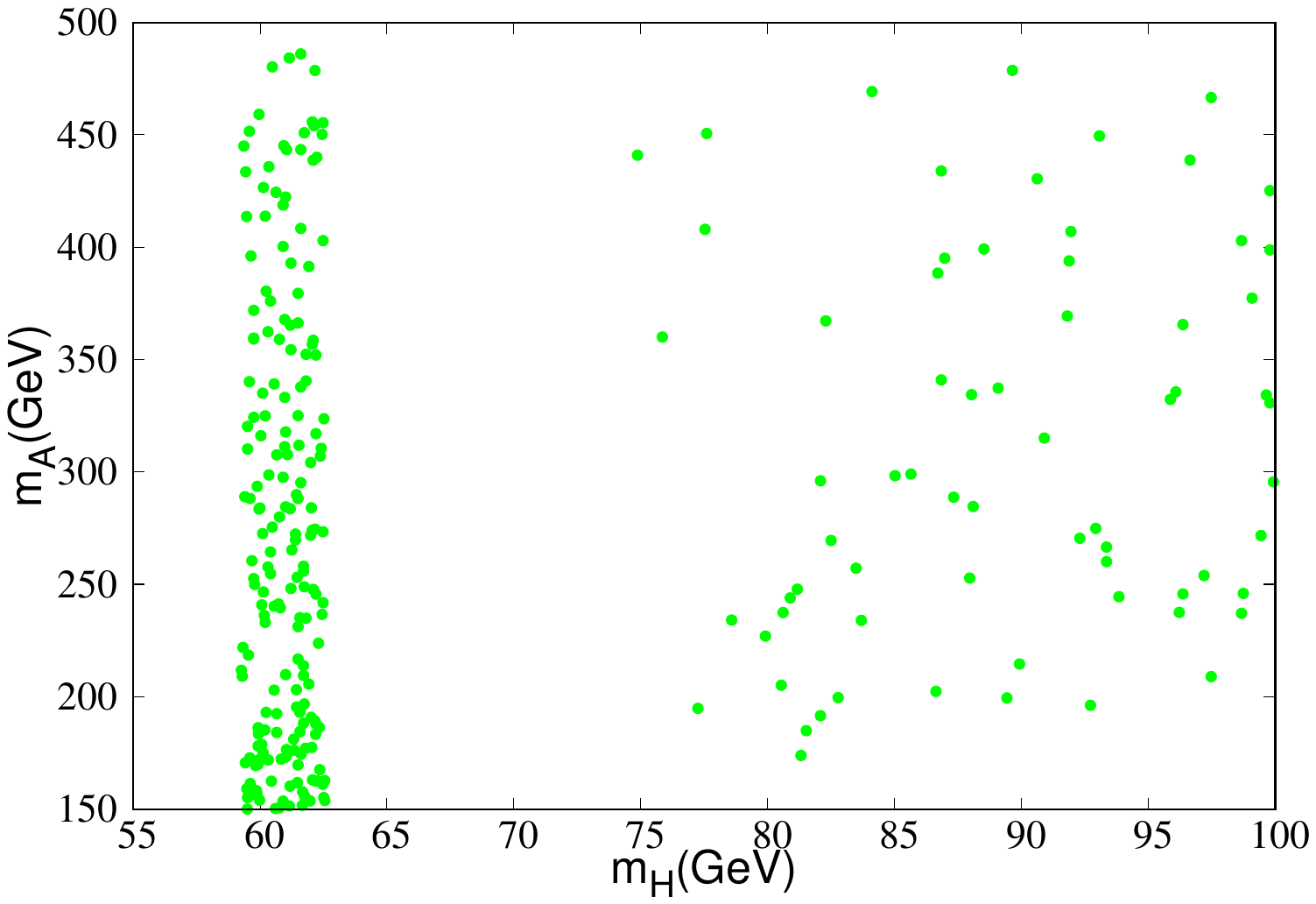}    
        \caption{The parameter space for the $Z$ + MET signal in the $\lb m_H, m_A\rb$ plane. In particular, DM constraints lead to the gap between 63 and 75 \GeV ~for the dark matter mass (see text for details). 
        The scale of $m_A$ stems from requiring mainly on-shell decays into the target signature. 
        } 
        \label{fig:dm_constraints}
        \end{center}
    \end{figure}

{In Fig.~\ref{fig:dm_constraints}, we show the allowed points in the parameter space of $\lb m_H, m_A\rb$ after imposing all constraints discussed in section \ref{sec:idm} and requiring nearly on-shell decays of $A$.} 
The heavier dark matter masses become strongly constrained. Especially, the region above the Higgs resonance up to $\sim 72$ GeV, appears to be highly disfavored for the range of $m_A$ considered here. 
Nevertheless, this region (above the Higgs resonance up to $\sim 72$ GeV) can be quite interesting from the point of DM phenomenology under certain circumstances.  When the mass differences between $H$ (DM) and $A$ or $H^{\pm}$ are small enough, coannihilation channels open up. We will discuss this in more detail below.

Now we perform recasting of the ATLAS search~\cite{ATLAS:2021gcn} in the $Z(\ell^+\ell^-)$+MET final state on the DM allowed parameter space shown in Fig.~\ref{fig:dm_constraints}. We emphasize that first we only focus on the lepton pair coming from on-shell $Z$. 

\begin{figure}[!hptb]
        \begin{center}
        \includegraphics[width=0.6\textwidth]{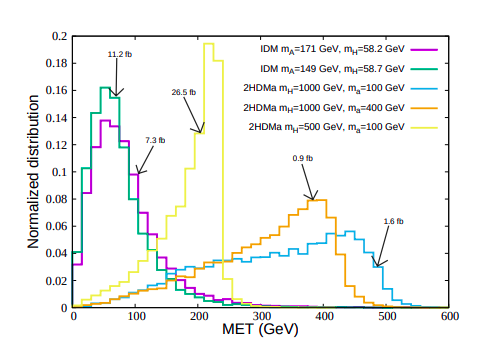}
        \caption{MET distribution for a few benchmarks in the two models 2HDMa and IDM. We also give production cross sections for the different scenarios. All 2HDMa points were excluded, and all IDM points allowed by the search considered here.
        }
        \label{fig:kinematics}
        \end{center}
    \end{figure}

We find that, when we consider only the leading production channel ($p p \to H A$), it does not impose any further constraint on the allowed parameter space. The reason behind that is as follows. The aforementioned search is optimized to constrain the benchmark DM scenario \thdma. The diagrams leading to the same final state, namely, $Z(\ell^+\ell^-$)+MET exhibit different production topologies, see figures \ref{fig:feynman_2hdma} and \ref{fig:feynman_IDM} for the \thdma{} and IDM, respectively.  This in turns has an effect on the kinematics of the final state particles. As an example, we present the distribution of MET in both models, a major kinematical observable used for analysis, in figure \ref{fig:kinematics}.

\begin{figure}[!hptb]
        \begin{center}
        \includegraphics[width=0.6\textwidth]{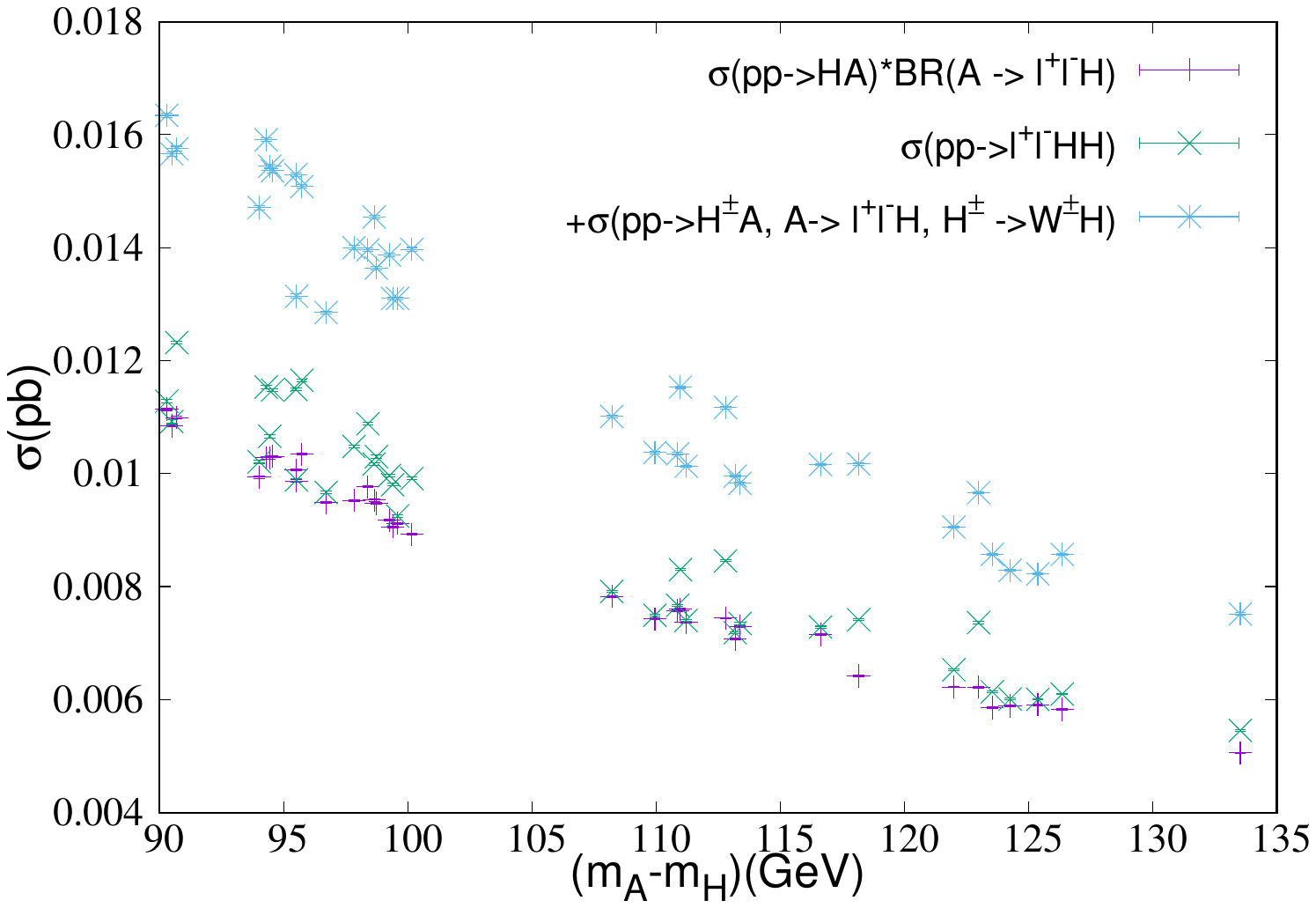}
        \caption{Production cross-section of $pp \rightarrow H A$ as a function of $m_A-m_H$, for the points used in Fig.~\ref{fig:crosssec}. Labelling of points is the same as in that figure.}
        \label{fig:mHmA}
        \end{center}
    \end{figure}

 It can be seen from this figure that all three chosen benchmarks of 2HDMa, peak at larger MET compared to all three IDM benchmarks. In order to understand this behavior, we compare the Feynman diagrams of the major production process in both models (Fig.~\ref{fig:feynman_2hdma} and \ref{fig:feynman_IDM}). In case of 2HDMa, a heavy on-shell scalar $H$ is being produced and then further decaying into $Z$ and $a$. It is expected in this case that both $Z$ and $a$ will have sufficient $p_T$, and subsequently large MET in the final state. 
On the other hand, in the IDM case, the regions of the parameter space with largest possible event rate, correspond to small mass difference between $A$ and $H$ (see Fig.~\ref{fig:mHmA}). 

\begin{figure}[!hptb]
        \begin{center}
        \includegraphics[width=0.45\textwidth]{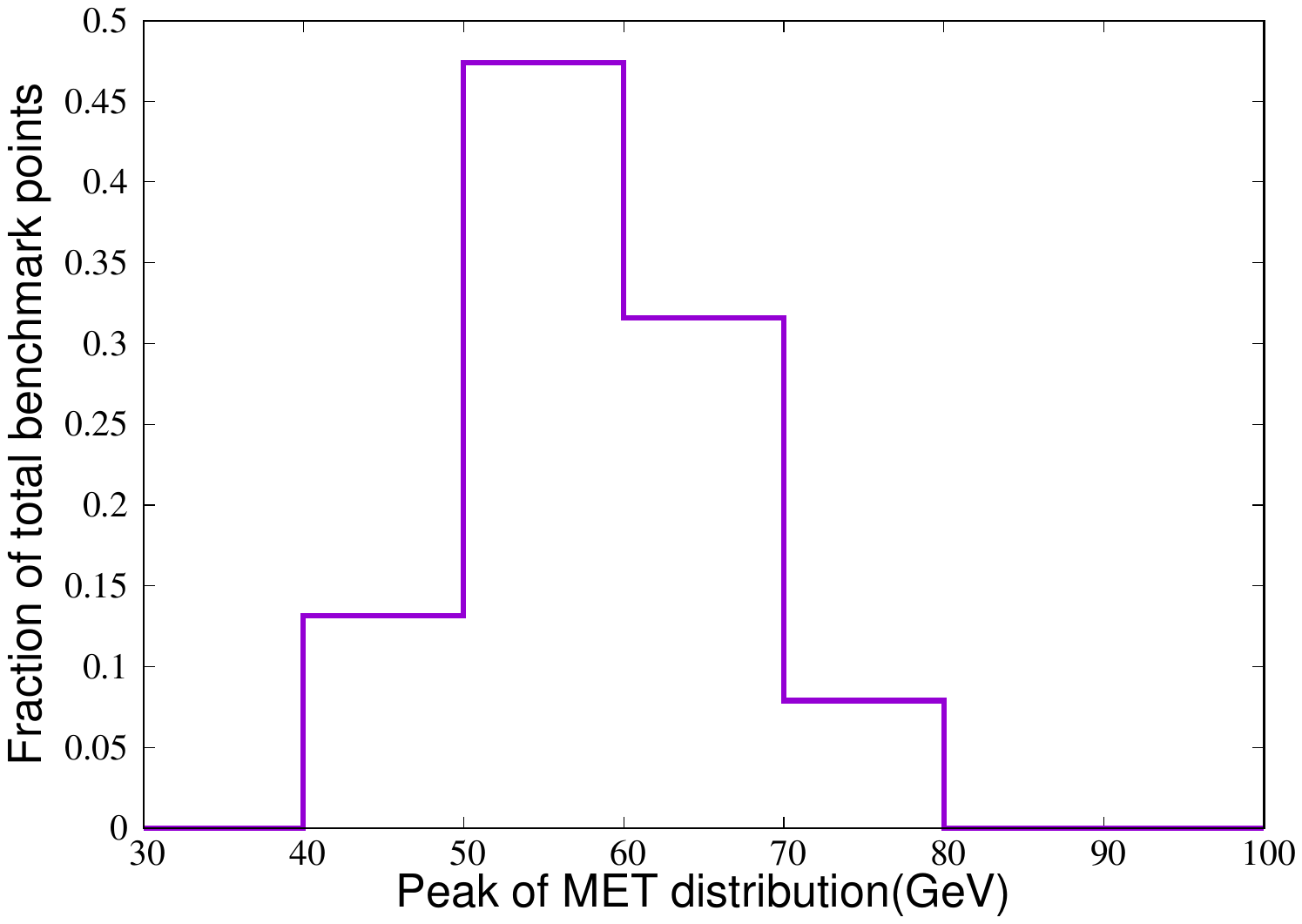}
        \includegraphics[width=0.45\textwidth]{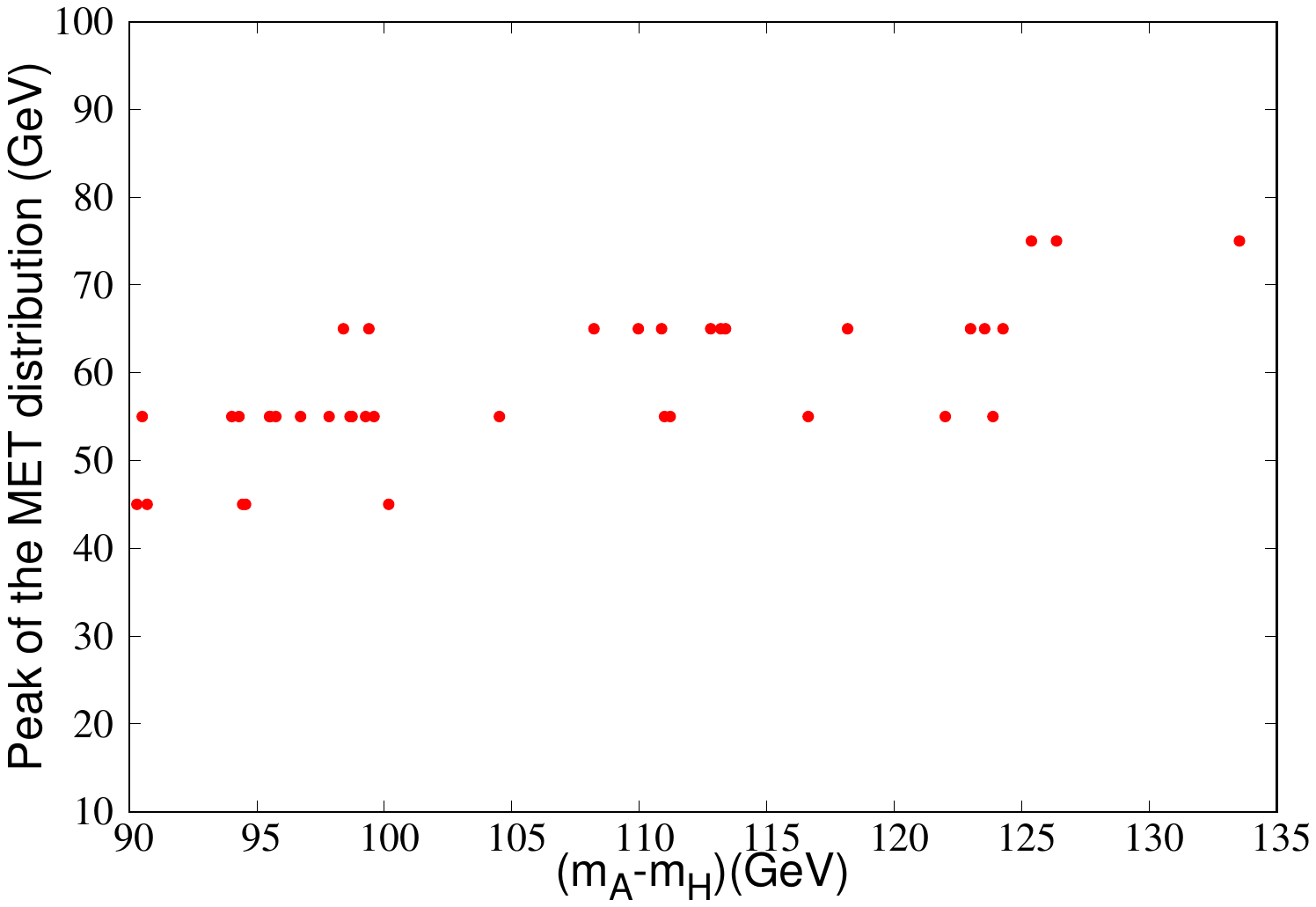}
        \caption{{\sl Left:} Peak (bin with the maximum number of events in Fig.~\ref{fig:kinematics}) of MET distribution for the benchmarks of Inert Doublet Model in Fig.~\ref{fig:crosssec}, considering $pp \rightarrow HA$ production only. {\sl Right:} the peak of the MET distribution as function of the mass difference $m_A - m_H$. 
        }
        \label{fig:metpeak}
        \end{center}
    \end{figure}

{{The on-shell $Z$ boson from the decay of $A$ has typically a $p_T$ $\sim 60$ GeV for our scenarios due to the relatively small mass difference $\sim 90$-$130\,\GeV$ between $A$ and $H$, which does not allow for a large phase space if the $Z$ boson is supposed to be on-shell. Consequently, the MET recoiling against the $Z$ is also typically very small. Figure~\ref{fig:metpeak} {\sl (left)} shows that the MET distribution peaks between 40-80 GeV. One can also see from Fig.~\ref{fig:metpeak} {\sl (right)} the peak shifts towards higher values with increase in the mass difference $m_A-m_H$.}} 

Therefore, the IDM benchmarks, even after having larger production rate compared to the \thdma{} benchmarks in most cases, are all allowed by the experimental data, as long as only the dominant contribution (on-shell $HA$ production) is taken into account.

\begin{figure}[!hptb]
        \begin{center}
        \includegraphics[width=0.48\textwidth]{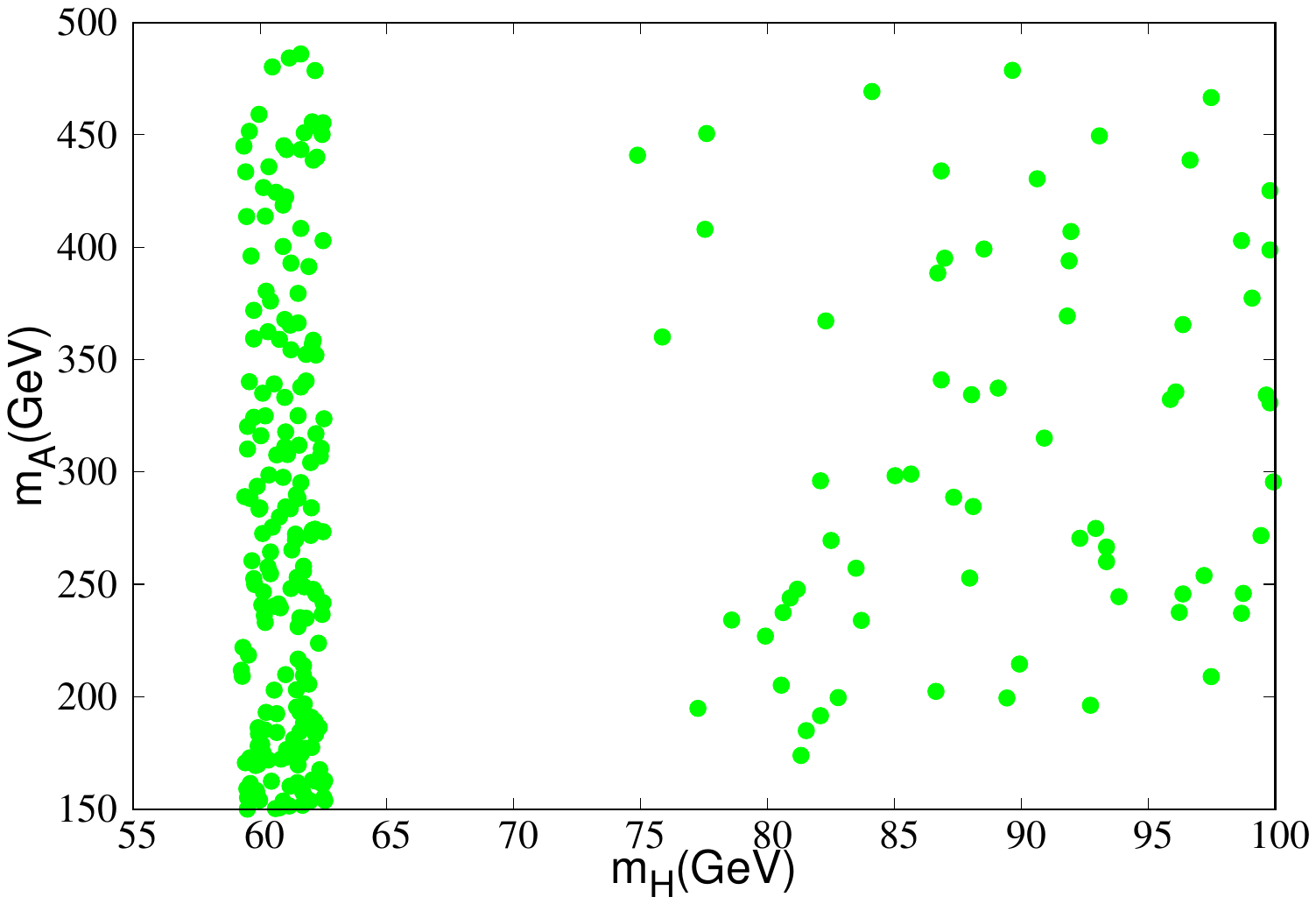}
        \includegraphics[width=0.48\textwidth]{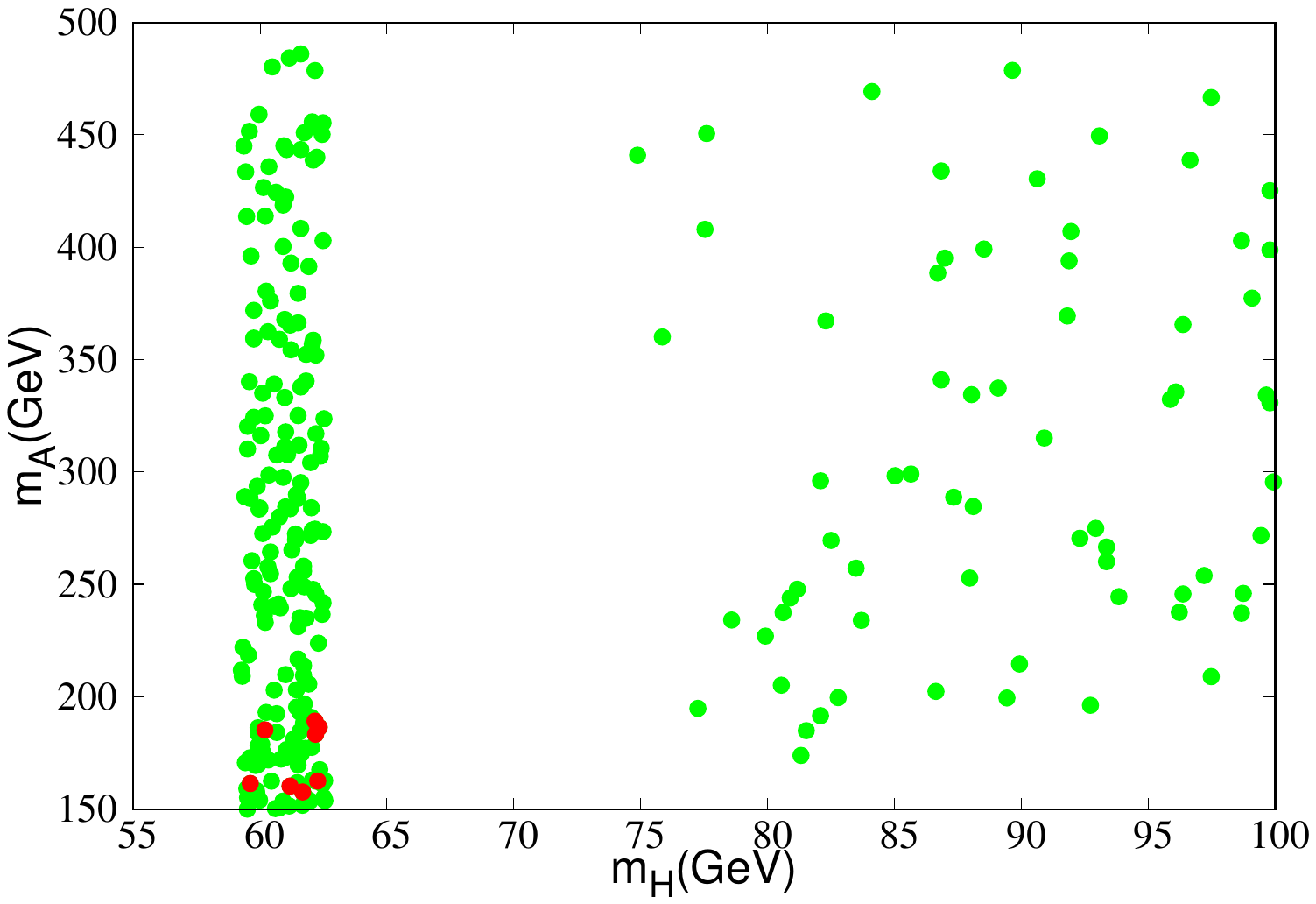} \\
         \includegraphics[width=0.48\textwidth]{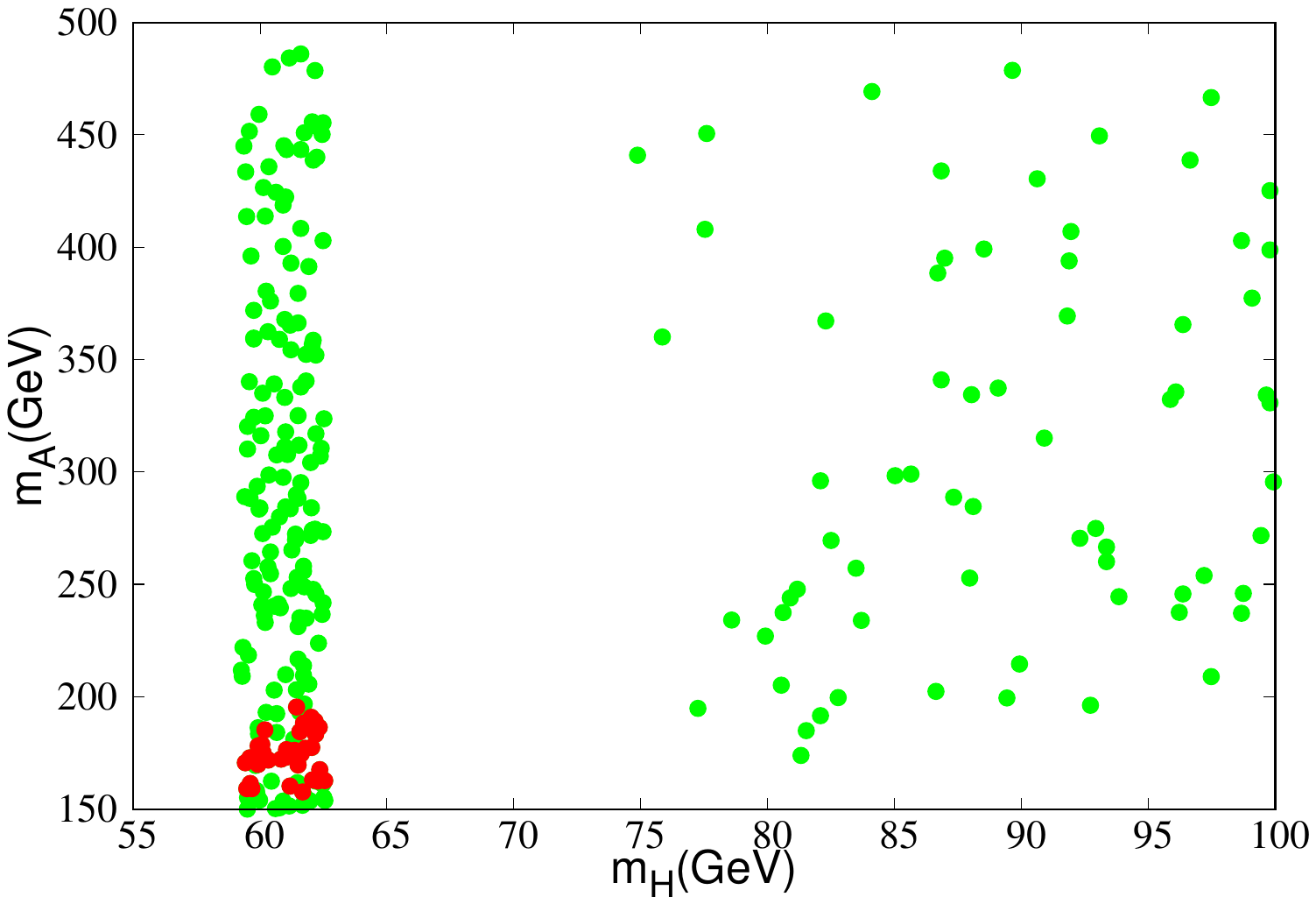}
      \caption{{\sl Top left:} The parameter space  allowed {\sl (green)} by recasting of ATLAS $Z$+MET search~\cite{ATLAS:2021gcn} when only on-shell $HA$ production is considered. {\sl Top right:} parameter space  allowed {\sl (green)} and excluded {\sl (red)} by recasting of ATLAS $Z$+MET search~\cite{ATLAS:2021gcn} when only on-shell $HA$ production and Higgs-Strahlung process are considered. {\sl Bottom:} parameter space allowed {\sl (green)} and excluded {\sl (red)} by recasting of ATLAS $Z$+MET search~\cite{ATLAS:2021gcn} when on-shell $HA$ production, Higgs-Strahlung and charged scalar associated production process are considered.
      }
        \label{fig:LHC_constraints}
        \end{center}
    \end{figure}

 Next we consider, in the IDM, the additional contribution from the sub-leading channels mentioned before. We show in Fig.~\ref{fig:LHC_constraints} {\sl (top right)}, that once the Higgs-Strahlung process is included, just a few points get excluded (shown in red). Furthermore, a broader region gets excluded when the events from charged Higgs associated production are also included in the signal final state (Fig.~\ref{fig:LHC_constraints} {\sl (bottom)}). It is worth noting that it is not really possible to get an exclusion contour in this case, as there are multiple parameters involved that lead to additional contributions from the respective sub-processes: $m_H, m_A, m_{H^{\pm}}$ as well as $\lambda_{345}$. In our analysis, we vary all of them simultaneously.
 
\begin{figure}[!hptb]
        \begin{center}
        \includegraphics[width=0.47\textwidth]{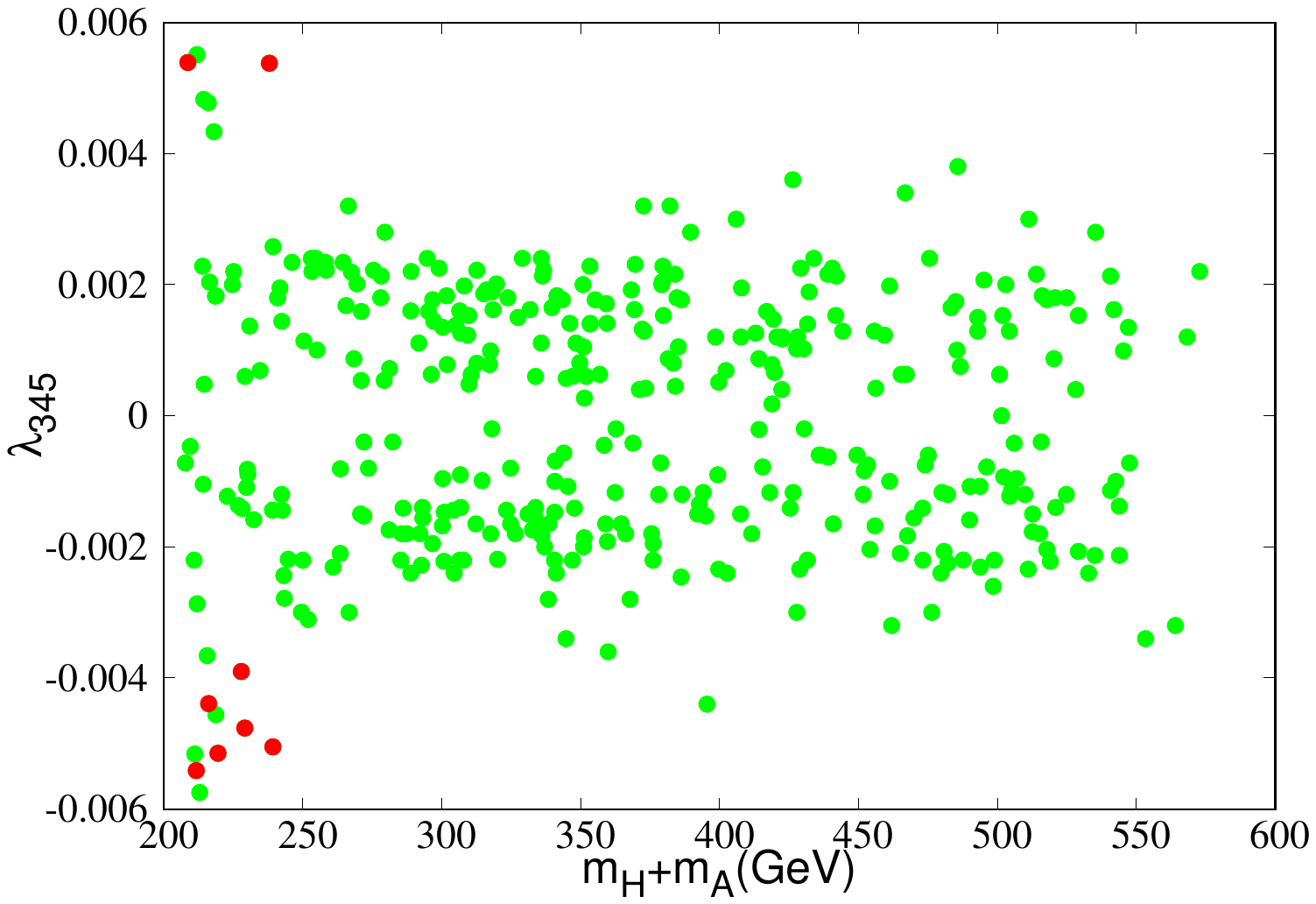}
            \includegraphics[width=0.47\textwidth]{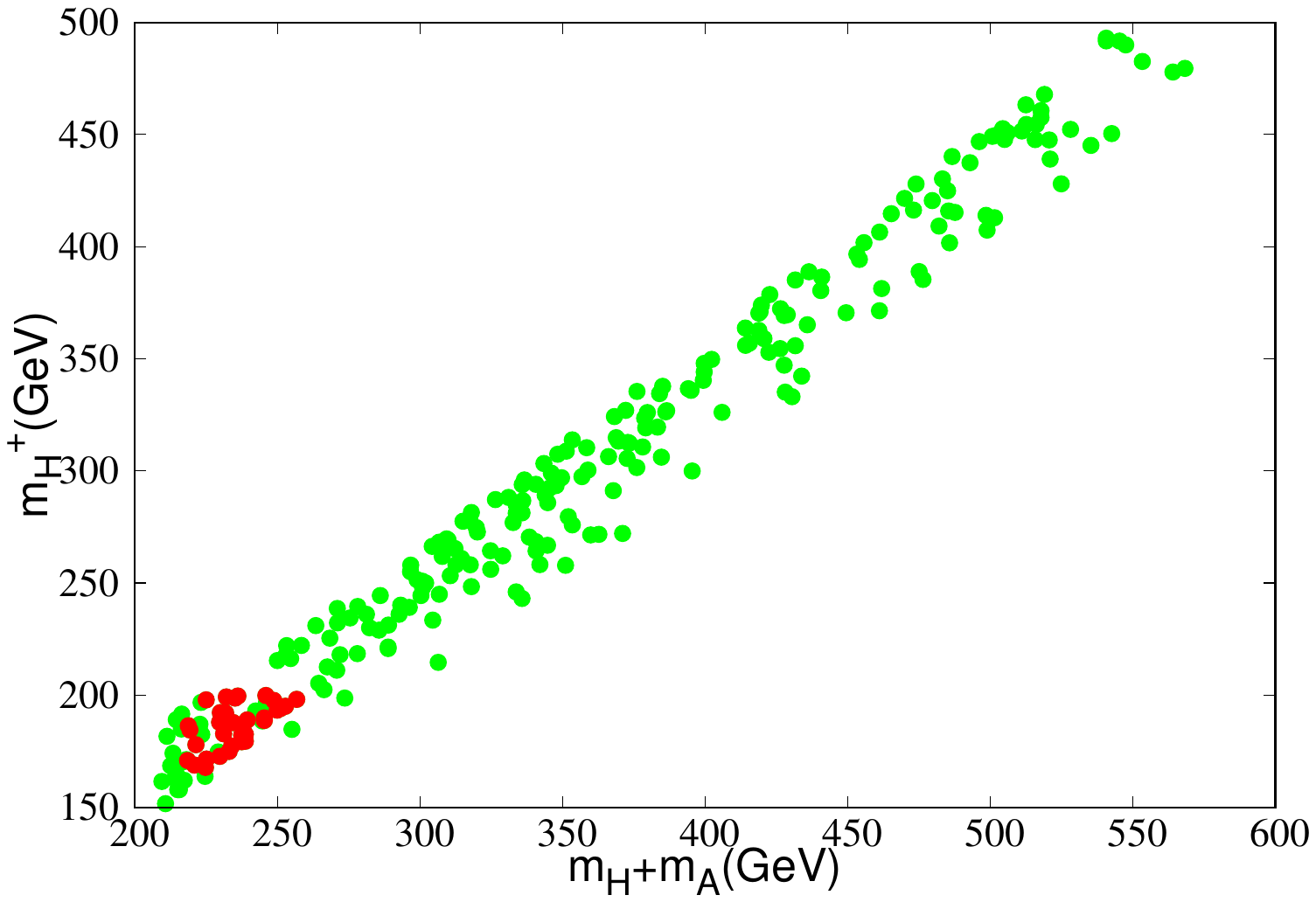}\\
            \includegraphics[width=0.47\textwidth]{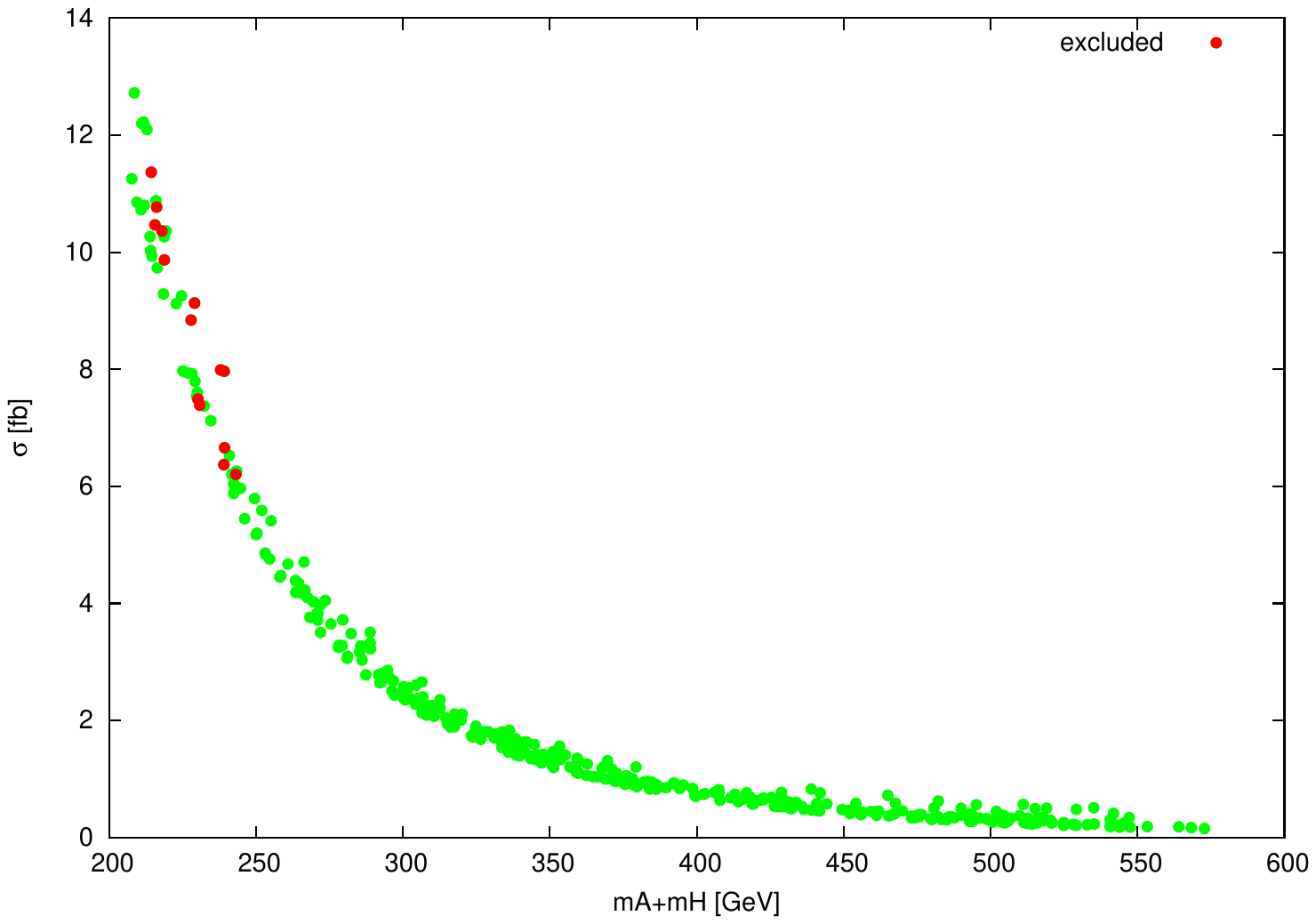}
        \caption{{\sl Top left:} Allowed parameter space {\sl (green)} and excluded region {\sl (red)} in $(m_A+m_H)-\lambda_{345}$ plane, after recasting when the $HA$-production and Higgs-Strahlung process is considered. {\sl Top right:} Allowed parameter space {\sl (green)} and excluded region {\sl (red)} in $(m_A+m_H)-m_{H^{\pm}}$ plane after recasting when $HA$-production, Higgs-strahlung and charged scalar associated production is considered. {\sl Bottom:} Allowed and excluded points with respect to the total production cross section.}
        \label{fig:lambda_charged}
        \end{center}
    \end{figure}

In order to see the impact of the other parameters, we show the allowed parameter space in the plane of $(m_H+m_A)-\lambda_{345}$ (in Fig.~\ref{fig:lambda_charged} {\sl (left)}) and $(m_H+m_A)-m_{H^{\pm}}$ (in Fig.~\ref{fig:lambda_charged} {\sl (right)}). We can see that regions that are sensitive to our recast typically display large absolute values of $\lambda_{345}$ as well as light charged scalar masses, as these regions lead to enhanced rates from the subleading processes in Eqs.~(\ref{eq:higgsstrahlung}) and (\ref{eq:ws}), respectively. Finally, in the bottom plot of the figure we show the excluded and allowed points with respect to the total production cross section. The minimal cross section we are sensitive to is 6~fb.
 
\begin{figure}[!hptb]
        \begin{center}
        \includegraphics[width=0.6\textwidth]{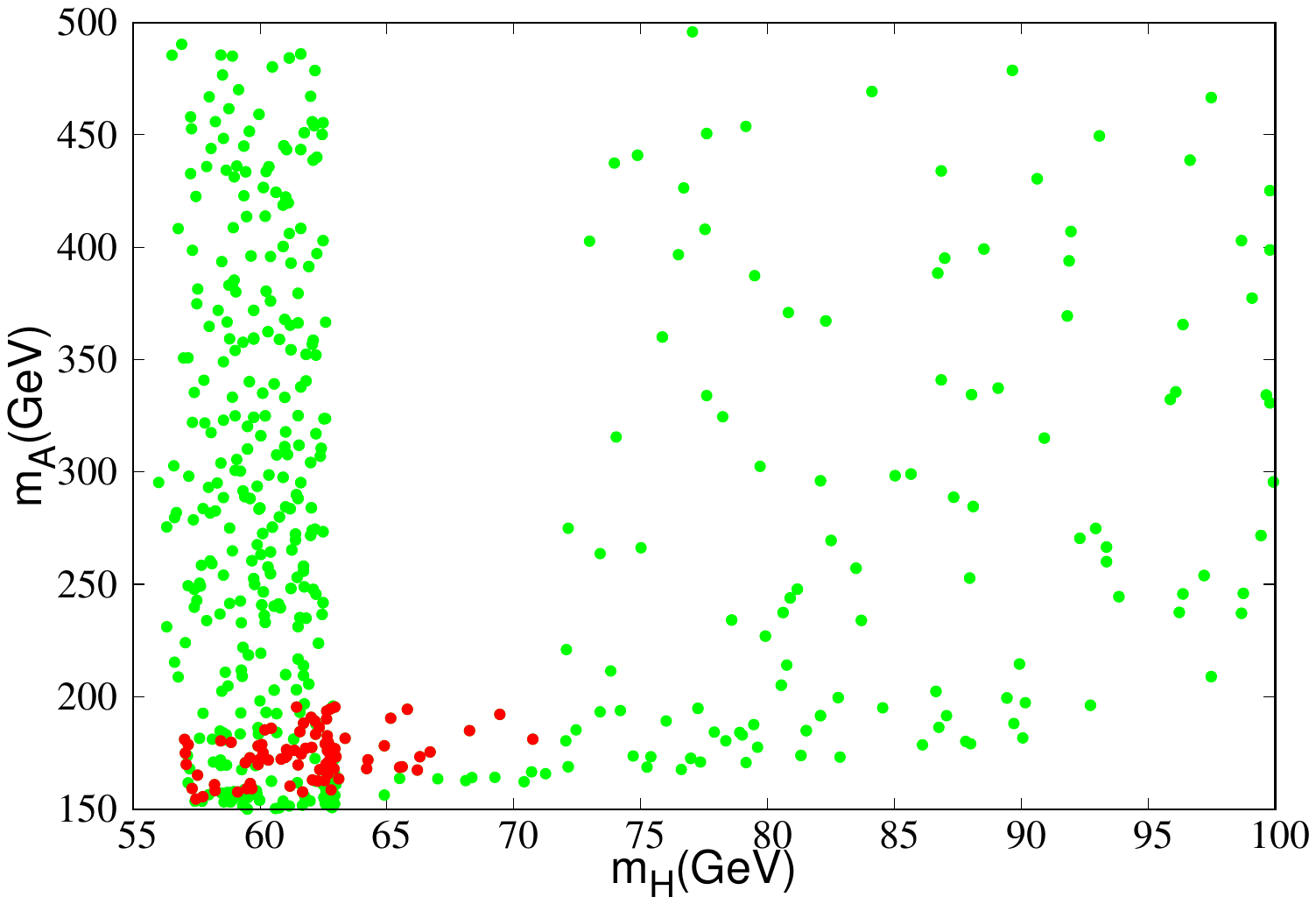}
        \caption{Allowed {\sl (green)} and excluded region {\sl (red)} when the full process is considered and no dark matter constraints are imposed.
        } 
        \label{fig:nodm}
        \end{center}
    \end{figure}

Finally, in order to see the full impact of the aforementioned $Z$+MET search at ATLAS, we also analysed a larger parameter space, without imposing the DM constraints and show the allowed and excluded region in Fig.~\ref{fig:nodm}.  We observe that when the dark matter constraints are not imposed, a larger region of parameter space opens up.  In particular, ignoring dark matter constraints opens up the region for dark matter masses $\lesssim\,60\,\GeV$ as well as allows for some additional points with masses $\lesssim\,80\,\GeV$.\footnote{Note that in fine-tuned scans more regions become available, see discussion below.} We further note that $150$~GeV $\lesssim  m_A \lesssim 200$ GeV and $55 \lesssim m_H \lesssim 70$~GeV region is strongly constrained by the recasting of the ATLAS  $Z$+MET search.

\begin{figure}[!hptb]
        \begin{center}
         \includegraphics[width=0.6\textwidth]{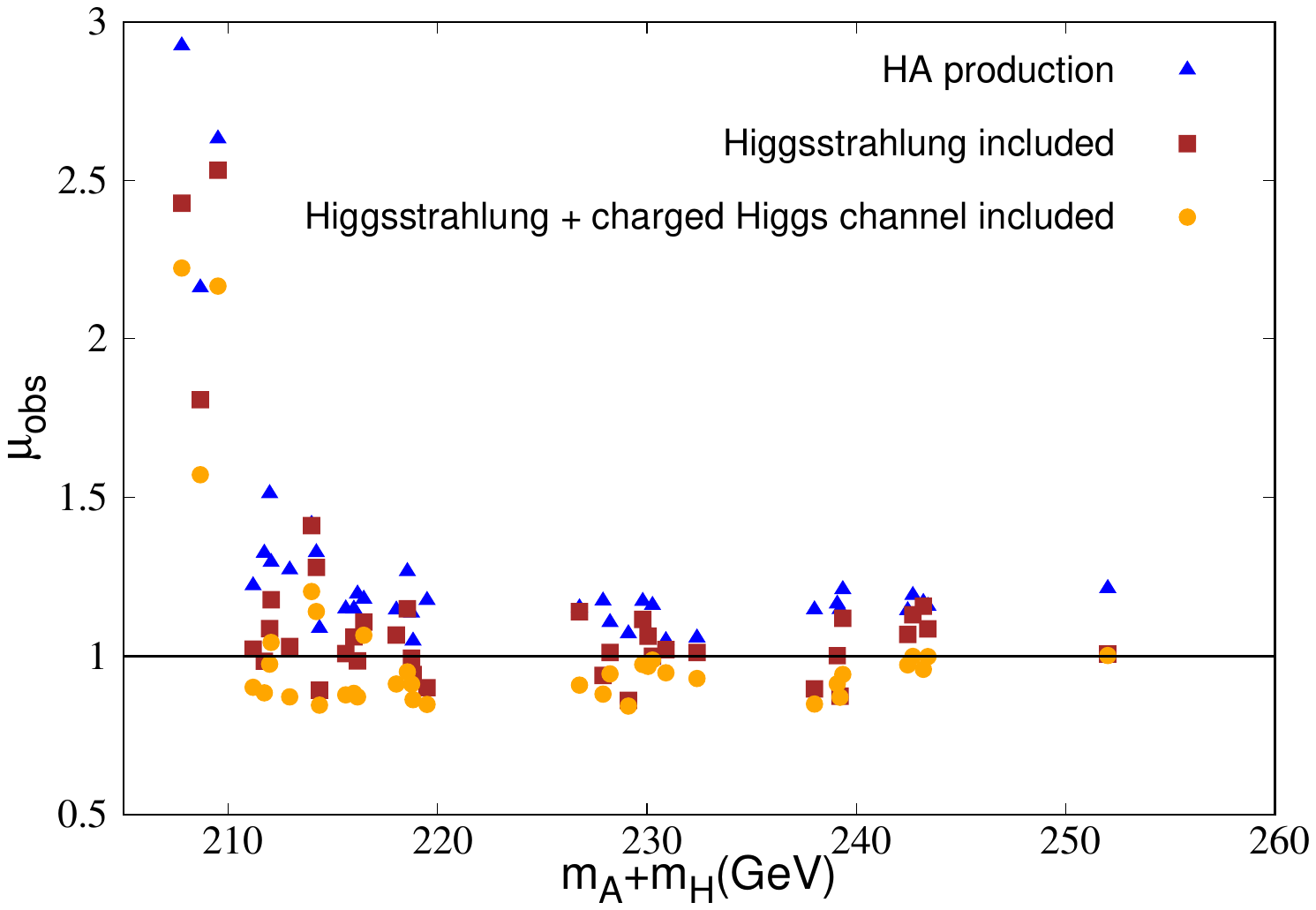}
        \caption{Representative  $\mu_{\text{obs}}$ values for leading contribution and as well as sub-leading contributions in IDM. Points with $\mu_{\text{obs}} < 1$ are excluded. 
        }
        \label{fig:r_multibin}
        \end{center}
    \end{figure}

We also display the distribution of $\mu_{\text{obs}}$, defined in Eq.~\eqref{eq:rdef} in Fig.~\ref{fig:r_multibin} for the representative points shown in Fig.~\ref{fig:crosssec}. The parameter points with $\mu_{\text{obs}} < 1$ are excluded by the experimental data. It is clear from this figure that, with leading contribution, the entire region considered is allowed (purple points). Once the sub-leading effects are included, a fraction of points becomes excluded (maroon and yellow points). We note that the exclusion is weak, ie.\ $\mu \gtrsim 0.9$. The $\sim 5\%$ theoretical uncertainty, evaluated in \mg, on the observed signal strength is due to PDF and scale variation. Additionally, 1-$\sigma$ experimental uncertainties on the expected signal strength are of the order 30-40\%. Therefore the total uncertainties by far exceed the exclusion margin for our benchmark points. This proves generally poor sensitivity of the search to \thdma{} scenarios.

Finally, we focus on the white region in Fig \ref{fig:dm_constraints}, for DM masses between 62--75 \GeV. In order to understand the characteristics of this part of the parameter space in more detail, we constrained ourselves to a region with $|\lam_{345}|\,\lesssim\,0.003$, and we see that only a tiny mass gap between $H$ and $A$ is allowed by the DM constraints, when $m_H \lesssim 70$ GeV. This is due to the fact that when the mass gap between the dark matter $H$, and the second lightest dark sector particle is small, coannihilation channels start to play a role. 
Therefore, it is possible to produce sufficient relic density to not overclose the universe with an extremely small portal coupling $\lambda_{345}$, thereby naturally satisfying the stringent direct detection bound. 

\begin{figure}[!hptb]
        \begin{center}
 \includegraphics[width=0.6\textwidth]{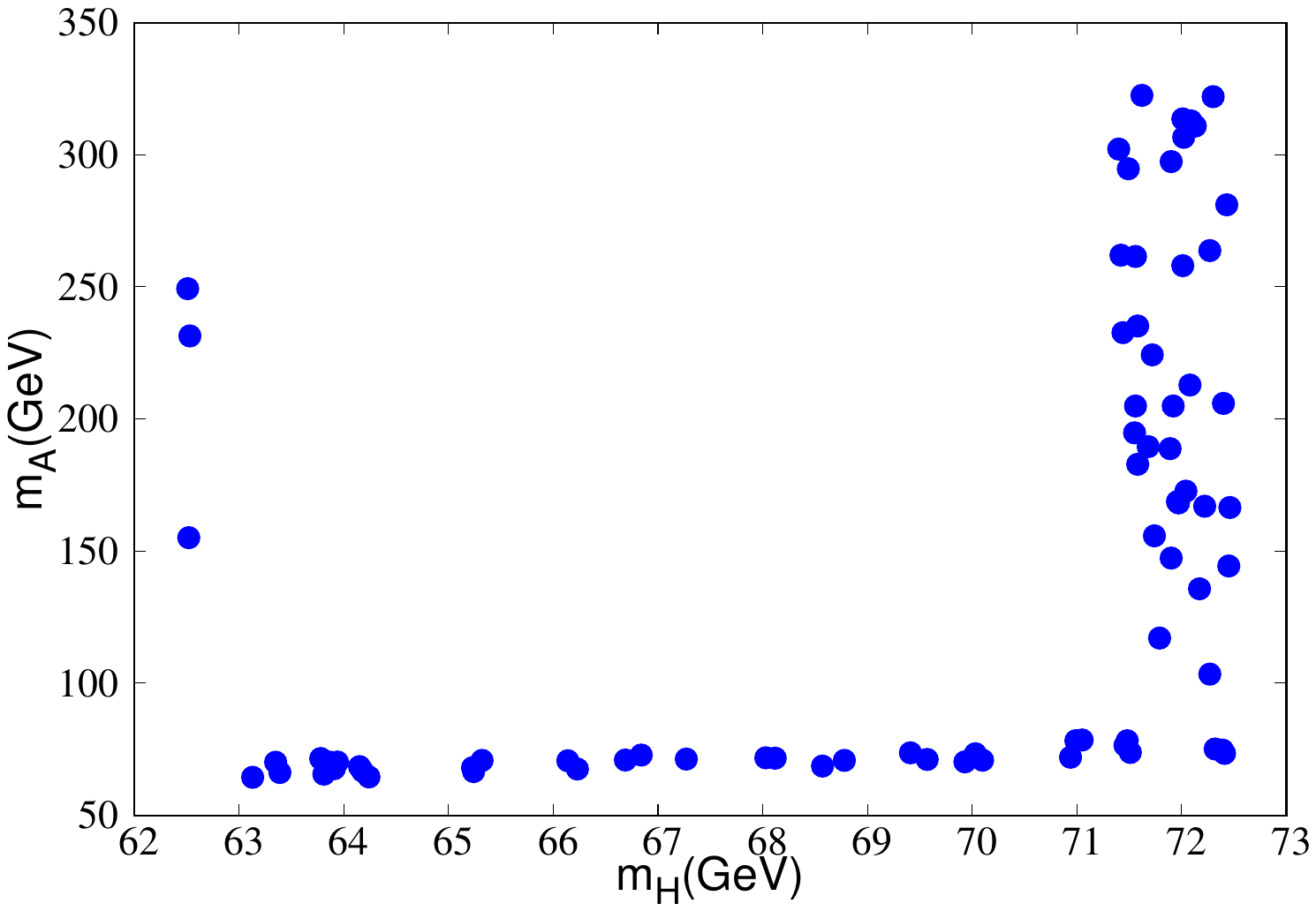}
 \caption{Allowed points from the dedicated scan with $|\lam_{345}|\,\lesssim\,0.003$. The allowed points between the $\frac{m_h}{2}$ and $\sim 72$ GeV correspond to $m_A-m_H < 10$ GeV. 
 } 
      \label{fig:small_massgap}
        \end{center}
    \end{figure}

We show this in Fig~\ref{fig:small_massgap}. It is clear that the mass-gap between $H$ and $A$ being small ($\lesssim 10$) GeV, such region cannot be constrained by the on-shell Z + MET search. We will probe this region in Section~\ref{sec:softleptons} using the soft-lepton search.

\subsection{VBF channel}

In this section, we discuss the results of recasting the search for $h\,\rightarrow\,\text{invisible}$ \cite{ATLAS:2022yvh} using full Run 2 data, that provide an update of the early Run 2 analysis presented in \cite{Dercks:2018wch}. Note that for this the relevant leading order parameters are given by $m_H$ and $\lam_{345}$. 

As in \cite{Dercks:2018wch}, in order to demonstrate the effects of the recast, we here loosen some of the constraints we introduced in section \ref{sec:idm}; in particular, we temporarily do no take constraints from dark matter observables as well as the inert vacuum condition into account, in order to allow for large $\lam_{345}$ values. We then apply the bounds of the above search on the remaining parameter space.

\begin{center}
     \begin{figure}[htb!]
        \begin{center}
 \includegraphics[width=0.7\textwidth]{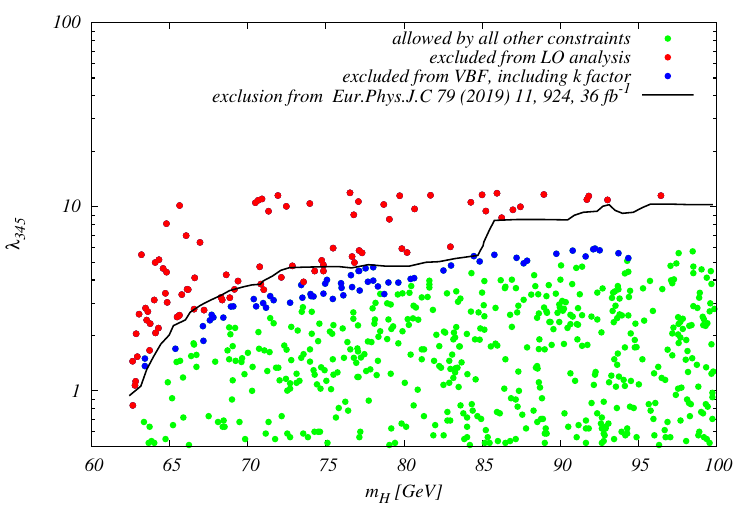}
 \caption{Exclusion from the VBF search for $h\,\rightarrow\,\text{invisible}$ \cite{ATLAS:2022yvh}. Shown are allowed and excluded points in the $\lb m_H,\,\lam_{345} \rb$ plane using LO as well as including an overall $K$-factor for the signal rate, see text for details. For reference, we also display the exclusion curve from earlier work \cite{Dercks:2018wch} using early Run II data. 
        \label{fig:vbf_exclusion_log}}
        \end{center}
    \end{figure}
\end{center}

The constraints on the IDM from VBF production mode are presented in Fig.~\ref{fig:vbf_exclusion_log}. We consider here the search for the invisible decays of the SM~\cite{ATLAS:2022yvh}. For masses $m_H\,\leq\,m_h/2$, one can easily apply the bounds on the branching ratio for this decay mode. We therefore here focus on dark matter masses that not kinematically allow for a direct decay $h \to H\, H$ having $m_H > m_h/2$. Therefore the constraints from Ref.~\cite{ATLAS:2022yvh} are applied to the kinematic setup where the SM Higgs boson is an off-shell mediator for production of two invisible $H$ bosons (or possibly other DM candidates).

The points in Fig.~\ref{fig:vbf_exclusion_log} are divided in 3 categories: the green points are allowed by the search~\cite{ATLAS:2022yvh} and all other constraints apart from the ones mentioned above; red points are excluded by the search ~\cite{ATLAS:2022yvh} assuming LO production cross sections, and the blue points are excluded assuming a global $K$-factor of 1.67 for the production corrected section, cf.\ Appendix \ref{sec:vbf_vali} and Ref.~\cite{Dercks:2018wch}. The solid line represents the exclusion bounds found in Ref.~\cite{Dercks:2018wch} using early Run 2 data (using $K$-factor and shape fit). As can be seen, the full Run 2 dataset improves somewhat the limits and sensitivity by a factor of 2 approximately. The range of excluded $\lambda_{345}$ couplings starts slightly below 1 near the threshold and remains in the range 2--3 for $m_H$ masses between 70 and 80~GeV.  
Note that, in contrast to the other channels discussed in this work, the search for $h\,\rightarrow\,\text{invisible}$ can easily be translated into other models that feature a scalar portal and coupling to (scalar) dark matter, as the search is determined only by two parameters, namely, the mass of the dark matter candidate as well as the coupling to the Higgs. Therefore, one can easily read off bounds for other scenarios from the above figure. This takes into account that the coupling of the 125 \GeV~ scalar to  other decay modes is not modified.

\subsection{Soft lepton channel\label{sec:softleptons}}

{In this subsection we take a look at a corner of parameter space with compressed mass spectra, $\Delta m \,\equiv\, m_A - m_H \sim \mathcal{O}(\mathrm{GeV})$. Such a small mass gap is interesting for DM phenomenology because such regions lead to co-annihilation of dark matter with another dark sector particle, namely $H$ and $A$ in our case. To derive constraints we recast the ATLAS SUSY search for electroweakinos~\cite{ATLAS:2019lng} with compressed mass spectra. The final state consists of two OSSF soft leptons (electrons or muons), missing transverse energy and a recoiling initial state radiation jet; see also Appendix~\ref{app:soft}. In the search for SUSY, such a final state appears mainly in the associated production of chargino and neutralino when a mass difference between neutralinos is small. In  the IDM, this final state  would appear in processes (a) and (c) depicted in Fig.~\ref{fig:feynman_IDM}. 
}

\begin{figure}
        \begin{center}
        \includegraphics[width=0.48\textwidth]{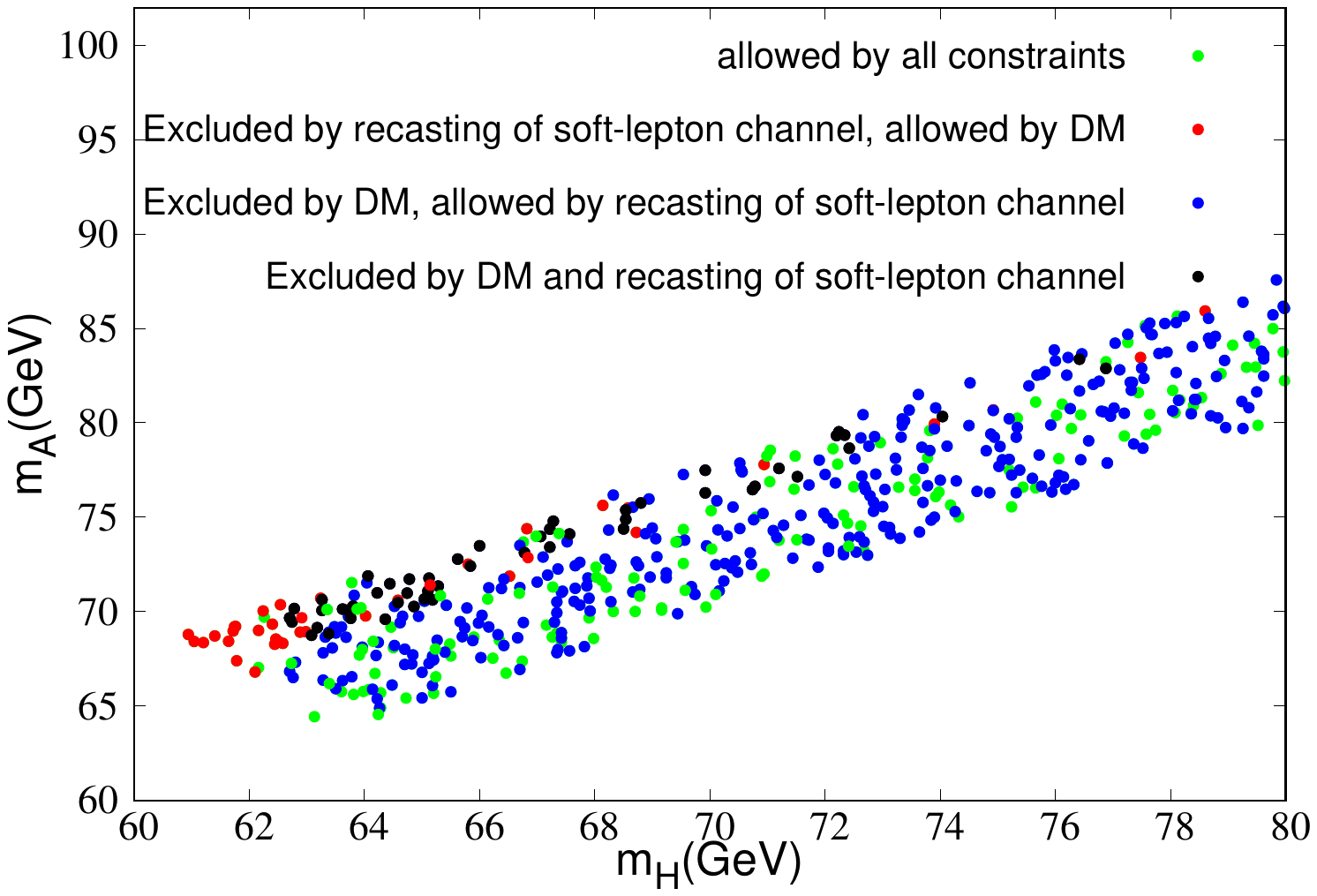}
 \includegraphics[width=0.48\textwidth]{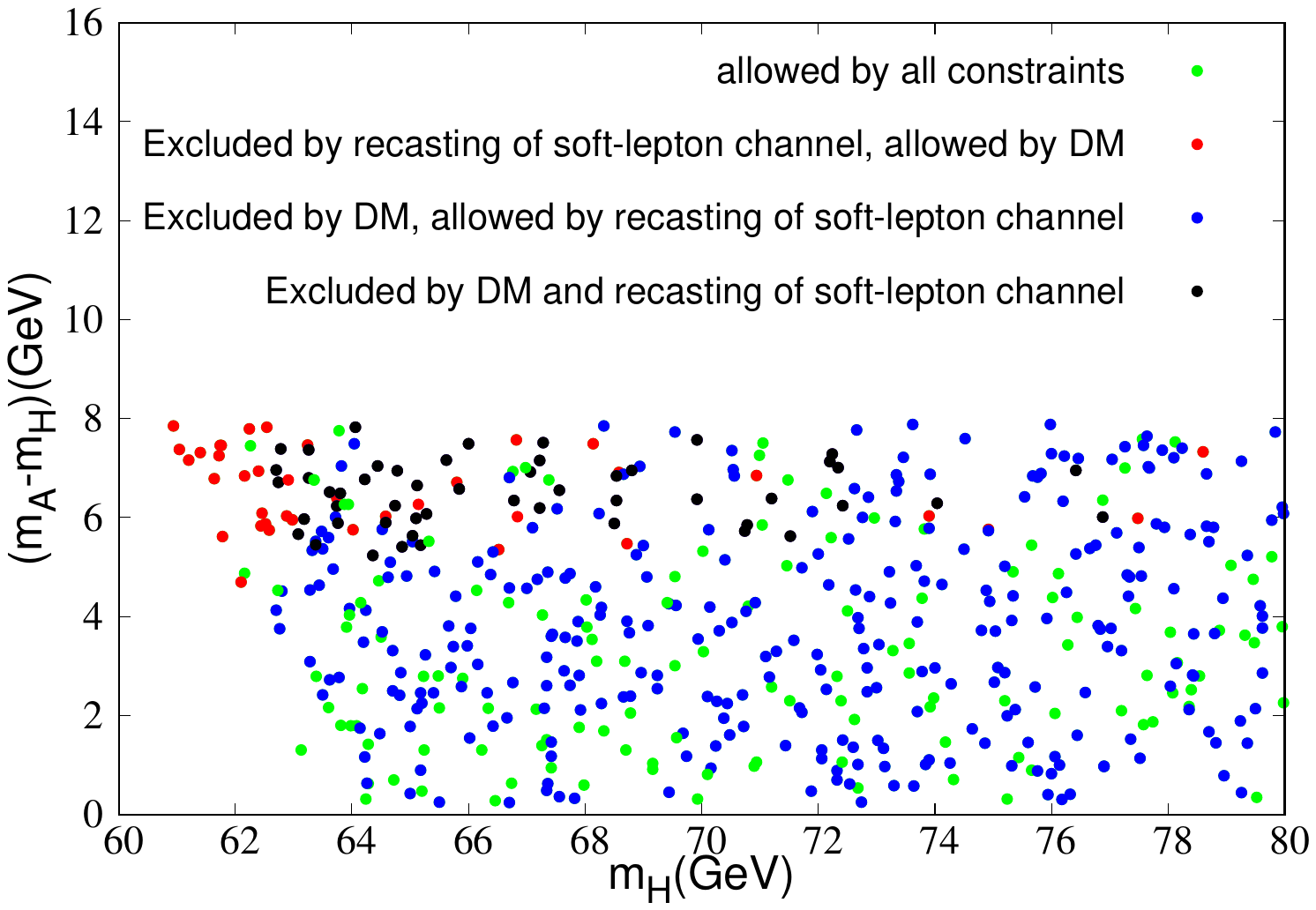}
 \end{center}
        \caption{The parameter space in $m_H$ vs $m_A$ {\sl (left)} and $m_H$ vs. $m_A - m_H$ {\sl (right)} plane. In both plots, the blue points are excluded by the dark matter constraints, but allowed by soft-lepton recast. The red points are allowed by dark matter constraints but excluded after recasting the soft lepton search~\cite{ATLAS:2019lng}. The black points are excluded by both dark matter constraints as well as soft lepton recast.
        }
        \label{fig:exclusion_soft}
        
    \end{figure}

The results of our analysis for the parameter points with the small mass difference, $\Delta m = m_A - m_H < 8$~GeV are displayed in Fig.~\ref{fig:exclusion_soft}. This region is of particular interest as it is still allowed by the recast of the SUSY search at DELPHI/LEP~\cite{EspiritoSanto:2003by} in terms of the IDM presented in \cite{Lundstrom:2008ai}.
We display the results in two panels: in the $m_H$--$m_A$ plane and in the $m_H$-$\Delta m$ plane. The search~\cite{ATLAS:2019lng} becomes sensitive for the points with $\Delta m \gtrsim 4.5\,\GeV$. Since there are several processes that contribute to the final state with two leptons, in particular the production of $A H^\pm$ pairs, one cannot draw an exclusion line in such a projection since an additional parameter, i.e.\ $m_{H^\pm}$ is also at play here. Still, the points with $\Delta m > 5\,\GeV$ and $m_H < 64\,\GeV$ are generally excluded by the soft lepton search. {It can be seen in Fig.~\ref{fig:exclusion_soft} that one cannot make an explicitly excluded region in the parameter space of $m_H$ and $m_A$: the black and red points excluded by the soft lepton search mix with blue and green points which are allowed by this search. This is a result of interplay between two production processes, as mentioned above, of $HA$ and $H^\pm A$. Nevertheless, this example shows that soft lepton SUSY searches can constrain other cosmologically interesting scenarios with small mass splittings between dark matter particles and other dark sector states. Dropping the DM constraints, the red and black points can be considered as a model-independent limit that can be obtained from collider searches in such experimentally challenging scenarios. Finally, we note that an interesting follow up of this analysis could be including and combining a similar soft lepton search from CMS~\cite{CMS:2021edw}. } 

\section{Summary}
\label{sec:summary}
In this work, we have confronted the Inert Doublet Model, a two Higgs doublet model with an exact $\mathbb{Z}_2$ symmetry, with various current LHC searches for final states that are also typical signatures of this model. So far, no dedicated LHC analysis is available that gives direct bounds on this scenario. We therefore rely on recasting several searches within the \CM{} framework. In particular, we discuss signatures containing dilepton and missing energy final states, VBF production of the SM-like Higgs boson decaying invisibly, as well as soft the lepton search with missing transverse energy.
We found that these searches rule out some regions of the IDM parameter space.

A particularly interesting point is the question whether searches that have the same final state signature but have been optimized for other models, for instance the 2HDMa in our case, and production topologies suffice to constrain the IDM. For this, we have made use of a search for the 2HDMa, a two Higgs doublet model with an additional pseudoscalar that serves as a mediator to dark matter in the gauge eigenstates. We found that the relatively high transverse missing energy cut render the current analysis quite insensitive to large regions of the IDM parameter space, despite higher cross sections. In fact, we found that including the dominant production process only does not lead to any exclusion, and that sub-leading contributions need to be taken into account.

Furthermore, we considered the VBF production of Higgs boson and its subsequent decay into DM pair. Although the limits from recasting the LHC results in this channel are in general less stringent compared to the DM constraints from direct searches, we would like to emphasize that this limit is model-independent and can be easily interpreted in terms of any model with Higgs-portal signature. 
Lastly, recasting the soft lepton search provides impressive limits on top of the already existing LEP bounds on the parameter points with compressed spectra of the DM particle and its partner, $H$ and $A$ in our context. 

Through this analysis, on the one hand we derive the most stringent limit on the IDM parameter space. On the other hand, we also provide a strong motivation for the experimental collaborations to design specific search strategies tailored towards the IDM. 
\section*{Acknowledgements}
TR and KR are supported by grant number HRZZ-IP-2022-10-2520 from the Croatian Science Foundation (HRZZ). The work of KR was supported by the OpenMAPP project via National Science Centre, Poland under CHIST-ERA programme (grant No.\ NCN 2022/04/Y/ST2/00186). TR and JL acknowledge support by the Deutsche Forschungsgemeinschaft (DFG, German Research Foundation) under Germany’s Excellence Strategy — EXC 2121 “Quantum Universe” — 390833306. TR also thanks the TH group at CERN for their hospitality while some of this work was completed.

\appendix
\section{Validation notes\label{app:validation}}

We provide here brief summary notes about the searches used in our study and their validation within the \CM{} framework. More details can be found in Ref.~\cite{Lara:2025cpm}.

\subsection{Mono-$Z$\label{app:1}}
\begin{center}
    \begin{figure}[htb!]
        \centering
        \includegraphics[width=0.7\textwidth]{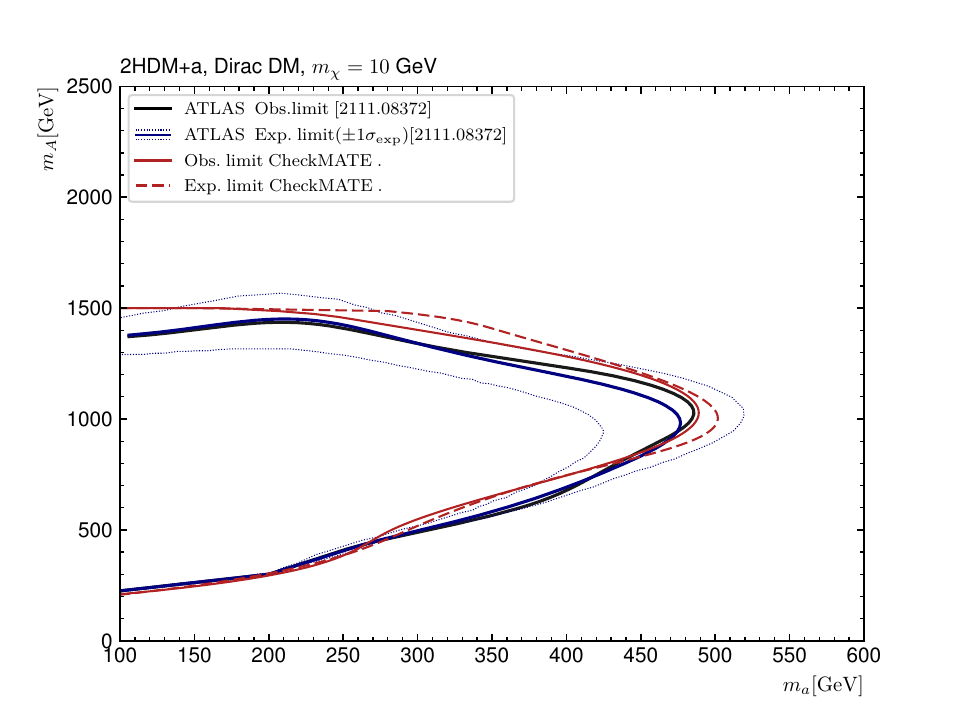}
        \caption{Validation plot of the HIGG-2018-26 search~\cite{ATLAS:2021gcn} for the \thdma{} model. Comparison inludes ATLAS observed limit (black line), ATLAS expected limit with experimental uncertainty (blue line), \CM{} observed limit (red solid line), and \CM{} expected limit (red dashed line). Courtesy of I\~naki Lara.
        \label{fig:ZMET_validation}}
    \end{figure}
\end{center}

This is a search \cite{ATLAS:2021gcn} (HIGG-2018-26) for invisible decays of a Higgs boson or dark matter particles produced in association with a $Z$ boson. The final state $Z$ boson is reconstructed from a same-flavour opposite-sign dilepton pair. One multibin signal region, optimized for the \thdma{} model~\cite{Bauer:2017ota}, is implemented in \CM{}. It is divided in 22 bins according to the transverse mass $m_\T$. The implementation uses a simplified likelihood model. The validation was performed for the \thdma{} assuming Dirac DM particle with $m_\chi = 10$~GeV and the comparison of ATLAS and \CM{} exclusion is shown in Fig.~\ref{fig:ZMET_validation}. There is a very good agreement both for observed and expected limits.

\subsection{VBF\label{sec:vbf_vali}}
\begin{figure}
    \centering
    \includegraphics[scale=0.5]{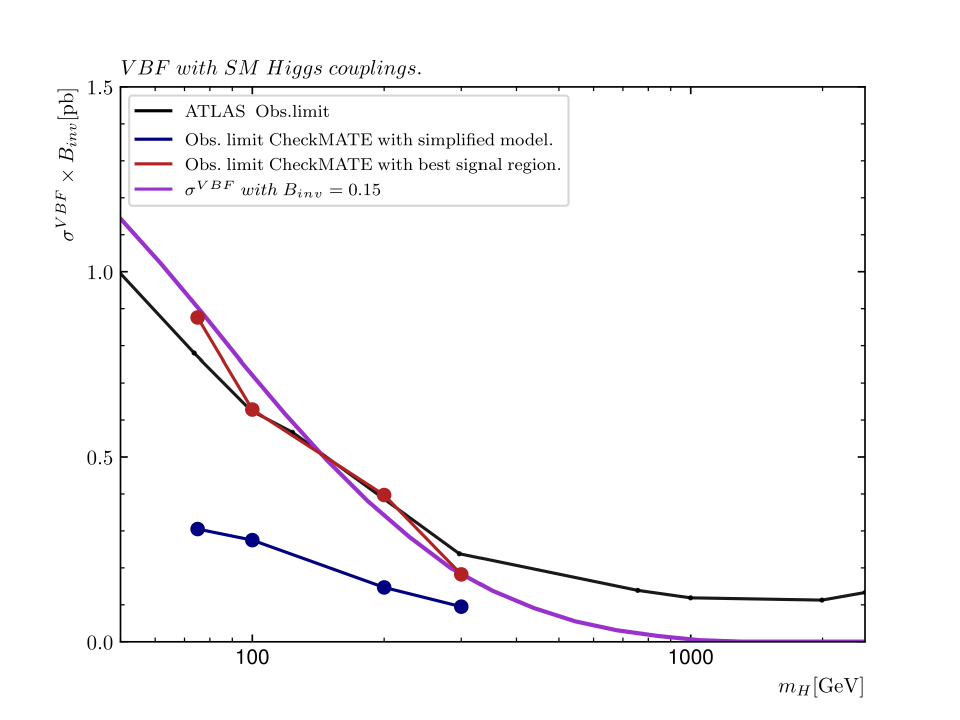}
    \caption{Comparison of observed limits on VBF Higgs boson production and invisible decay by ATLAS (black line) Ref.~\cite{ATLAS:2022yvh} and obtained using \CM{} from the best signal region method (red line) and the simplified likelihood shape-fit (blue line).
    \label{fig:vbfvalidation}}
\end{figure}

EXOT-2020-11 is a search \cite{ATLAS:2022yvh} for invisible decays of a Higgs boson produced in  vector boson fusion.  The final state consists of two forward jets consistent with a vector boson signature and missing transverse momentum. There is one multibin signal region divided into 16 bins according to a number of jets, invariant mass of the leading jets pair $m_{jj}$ and their angular separation. The validation is performed using a sample of events generated with \textsc{Powheg-Box}~\cite{Nason:2004rx,Frixione:2007vw,Alioli:2010xd} at NLO and hadronized with \textsc{Pythia}. For validation we compare limits on the VBF production and invisible decay (BR = 15\%) of a scalar particle with the SM Higgs boson couplings. Since there is no likelihood model provided we have used the simplified multibin shape fit~\cite{ATL-PHYS-PUB-2021-038} and best signal region method. We compare the results to the ATLAS observed limit in Fig.~\ref{fig:vbfvalidation}.   As can be seen, the shape fit result (blue line)) gives too strong constraints, while the best signal region method (red line) accurately reproduces ATLAS limit (black line). Therefore, in deriving constraints we only use the best signal region approach.

Furthermore, as noted in Ref.~\cite{Dercks:2018wch}, the above limits include higher-order corrections implemented in \textsc{Powheg}. On the other hand in the present study the IDM events are simulated in \mg{} at the leading order. To take this difference into account we introduce a $K$-factor, following a prescription of Ref.~\cite{Dercks:2018wch}. We simulate a sample of events for a 125~GeV invisibly decaying scalar and compare the number of events recorded in the signal region to the number obtained using \textsc{Powheg} in the validation procedure. This results in a value $K = 1.67$, which is in a very good agreement with the result of Ref.~\cite{Dercks:2018wch}. {Note, however, that this $K$-factor is derived for on-shell Higgs production, whereas in our case it is applied to off-shell scenarios.}  In the derivation of exclusion limits we compare the results with and without the $K$-factor.

\subsection{Soft leptons \label{app:soft}}

\begin{figure}
    \centering
    \includegraphics[scale=0.5]{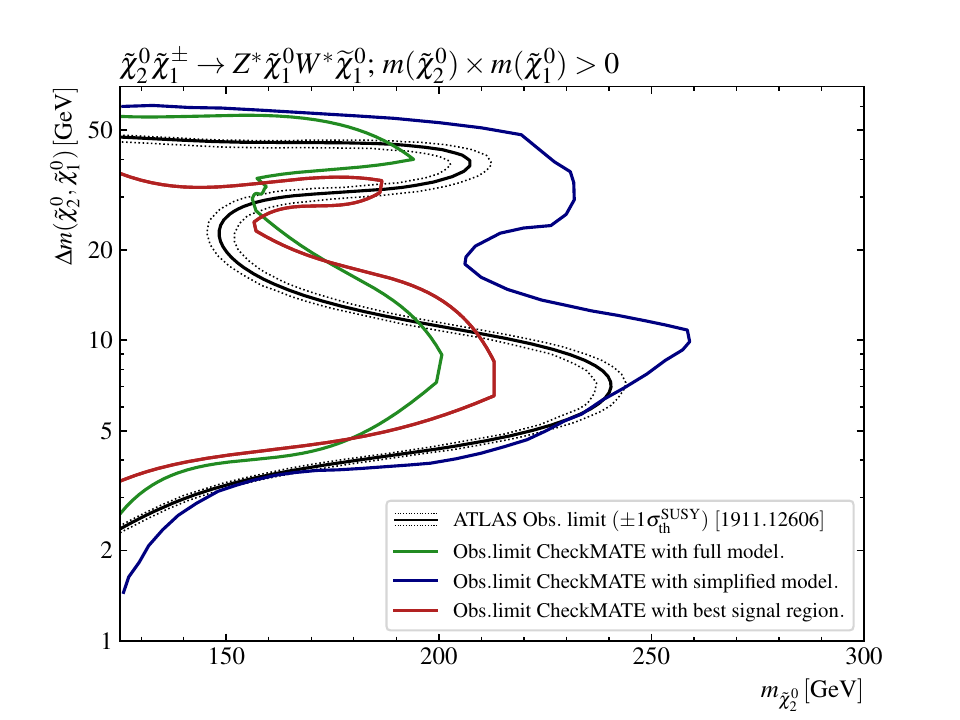}
    \includegraphics[scale=0.5]{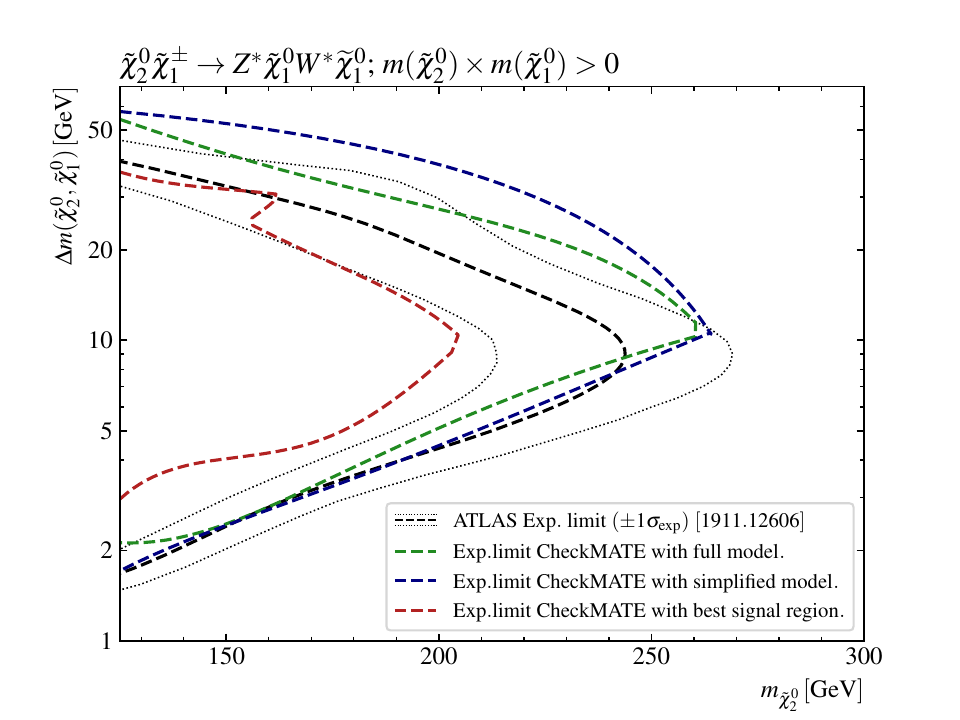}\\
    \includegraphics[scale=0.5]{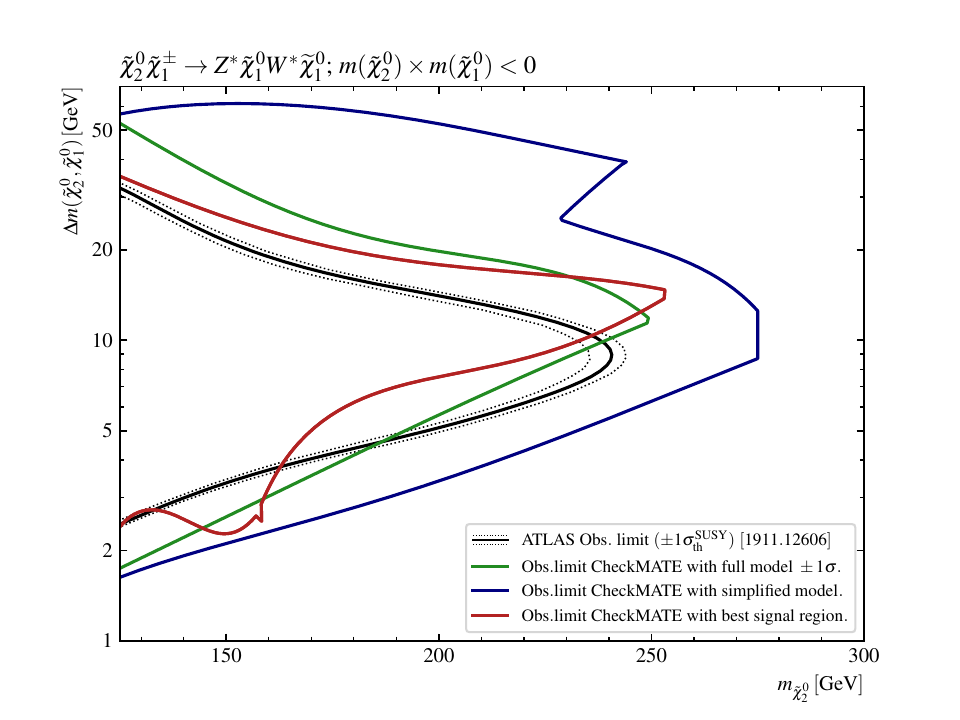}
    \includegraphics[scale=0.5]{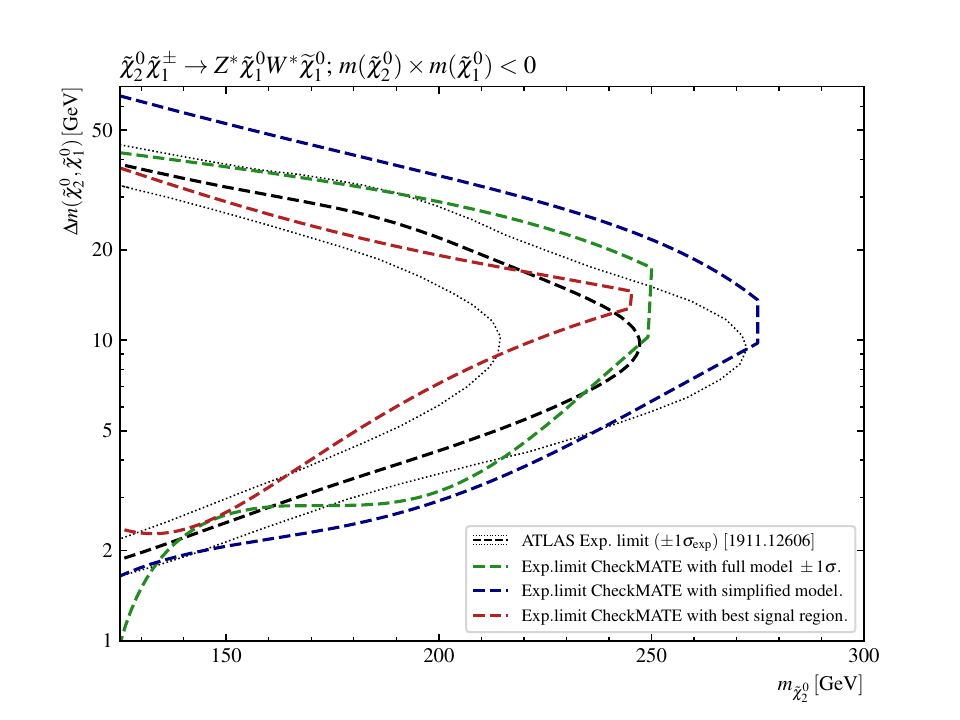}
    \caption{Comparison of limits between \CM{} and ATLAS on the production of wino-like $\tilde{\chi}^\pm_1 \tilde{\chi}^0_2$ pair, based on Ref.~\cite{ATLAS:2019lng}. Left column: observed limits; right column: expected limits; upper row: $m(\tilde{\chi}_{2}^{0}) \times m(\tilde{\chi}_{1}^{0})>0$ -- the same CP parity of neutralinos $\tilde{\chi}_{2}^{0}$ and $\tilde{\chi}_{1}^{0}$; lower row: $m(\tilde{\chi}_{2}^{0}) \times m(\tilde{\chi}_{1}^{0})<0$ -- the opposite parity. The convention which allows formally negative neutralino mass eigenvalues, while keeping mixing matrix real, was used here; see eg.\ Ref.~\cite{Choi:2005gt} for a discussion of CP properties of the neutralino sector and consequences for the decay kinematics. The plots compare three methods of deriving limits in \CM{}: best signal region (red), simplified statistical model (blue), and full statistical model (red). The full model is used to derive constraints on the IDM. }
    \label{fig:validation_soft}
\end{figure}

This is a search \cite{ATLAS:2019lng} for production of electroweakinos and sleptons in scenarios with compressed mass spectra. The final states contain two low $p_T$ leptons (opposite sign and same or different flavour) as well as missing transverse energy, $\met >  120$~GeV. The sensitivity of the search relies on additional initial state radiation jets which give transverse boost to the final state particles and adds missing transverse momentum. There are two multibin signal regions implemented in \CM{}:\footnote{The VBF SRs are currently not available.} \texttt{SR-EWK} is targeting the production of electroweakinos and divided into 44 bins according to the lepton pair invariant mass, \met, and lepton flavour; \texttt{SR-S} is targeting the production of sleptons and divided into 32 bins. The full likelihood model is provided. The validation plots for \texttt{SR-EWK}, which is relevant for the present study, are shown in Fig.~\ref{fig:validation_soft}. Throughout this work we use the full likelihood model for deriving constraints. 

\clearpage
\bibliography{lit.bib,lit_2.bib}
\bibliographystyle{utphys28mod}
\end{document}